\input epsf.tex
\documentclass[12pt]{article}
\usepackage{graphicx}
\usepackage{psfig,picinpar,floatflt,amssymb,epsf}
\renewcommand{\theequation}{\thesection.\arabic{equation}}
\csname @addtoreset\endcsname{equation}{section}

\def\bseq{\begin{subequation}}  % = 1a 1b
\def\eseq{\end{subequation}}
\def\bsea{\begin{subeqnarray}}  % = 1.1a 1.1b
\def\esea{\end{subeqnarray}}
\def\Tilde#1{\widetilde{#1}}                    % big tilde

\newcommand{\non}{\nonumber}

\newcommand{\beq}{\begin{equation}}
\newcommand{\eeq}{\end{equation}}
\newcommand{\bea}{\begin{eqnarray}}
\newcommand{\eea}{\end{eqnarray}}
\newcommand{\ena}{\end{eqnarray}}

\renewcommand{\a}{\alpha}
\renewcommand{\b}{\beta}

\renewcommand{\d}{\delta}

\newcommand{\pa}{\partial}
\newcommand{\g}{\gamma}
\newcommand{\G}{\Gamma}

\newcommand{\D}{\Delta}
\newcommand{\e}{\epsilon}

\renewcommand{\l}{\lambda}

\newcommand{\m}{\mu}

\newcommand{\n}{\nu}

\renewcommand{\S}{\Sigma}

\renewcommand{\o}{\omega}

\newcommand{\Db}{\bar{D}}

\newcommand{\Phib}{\bar{\Phi}}

\newcommand{\ad}{{\dot{\alpha}}}

\def\be{\begin{equation}}
\def\ee{\end{equation}}
\def\bea{\begin{eqnarray}}
\def\eea{\end{eqnarray}}
\def\ba{\begin{array}}
\def\ea{\end{array}}
\def\bi{\begin{itemize}}
\def\ei{\end{itemize}}

\def\Tr{{\rm Tr}}
\catcode`\@=11
\def\@citex[#1]#2{%
\if@filesw \immediate \write \@auxout {\string \citation {#2}}\fi
\@tempcntb\m@ne \let\@h@ld\relax \def\@citea{}%
\@cite{%
  \@for \@citeb:=#2\do {%
    \@ifundefined {b@\@citeb}%
      {\@h@ld\@citea\@tempcntb\m@ne{\bf ?}%
      \@warning {Citation `\@citeb ' on page \thepage \space undefined}}%
      {\@tempcnta\@tempcntb \advance\@tempcnta\@ne%
      \@tempcntb\number\csname b@\@citeb \endcsname \relax%
      \ifnum\@tempcnta=\@tempcntb %Number follows previous--hold on to it
        \ifx\@h@ld\relax%
          \edef \@h@ld{\@citea\csname b@\@citeb\endcsname}%
        \else%
          \edef\@h@ld{\ifmmode{-}\else--\fi\csname b@\@citeb\endcsname}%
        \fi%
      \else%   %  non-successor--dump what's held and do this one
        \@h@ld\@citea\csname b@\@citeb \endcsname%
        \let\@h@ld\relax%
      \fi}%
    \def\@citea{,\penalty\@highpenalty\,}%
  }\@h@ld
}{#1}}

\def\@citeb#1#2{{[#1]\if@tempswa , #2\fi}}
\def\@citeu#1#2{{$^{#1}$\if@tempswa , #2\fi }}
\def\@citep#1#2{{#1\if@tempswa , #2\fi}}

\def\bcites{         % cite with []'s
        \catcode`\@=11
        \let\@cite=\@citeb
        \catcode`\@=12
}

\def\upcites{         % cite with exponents
        \catcode`\@=11
        \let\@cite=\@citeu
        \catcode`\@=12
}

\def\plaincites{      % cite without brackets
        \catcode`\@=11
        \let\@cite=\@citep
        \catcode`\@=12
}

\newcount\hour
\newcount\minute
\newtoks\amorpm
\hour=\time\divide\hour by 60
\minute=\time{\multiply\hour by 60 \global\advance\minute by-\hour}
\edef\standardtime{{\ifnum\hour<12 \global\amorpm={am}%
        \else\global\amorpm={pm}\advance\hour by-12 \fi
        \ifnum\hour=0 \hour=12 \fi
        \number\hour:\ifnum\minute<10 0\fi\number\minute\the\amorpm}}
\edef\militarytime{\number\hour:\ifnum\minute<10 0\fi\number\minute}

\def\draftlabel#1{{\@bsphack\if@filesw {\let\thepage\relax
   \xdef\@gtempa{\write\@auxout{\string
      \newlabel{#1}{{\@currentlabel}{\thepage}}}}}\@gtempa
   \if@nobreak \ifvmode\nobreak\fi\fi\fi\@esphack}
        \gdef\@eqnlabel{#1}}
\def\@eqnlabel{}
\def\@vacuum{}
\def\marginnote#1{}
\def\draftmarginnote#1{\marginpar{\raggedright\scriptsize\tt#1}}
\def\draft{
        \pagestyle{plain}
        \overfullrule=2pt
        \oddsidemargin -.5truein
        \def\@oddhead{\sl \phantom{\today\quad\militarytime} \hfil
        \smash{\Large\sl DRAFT} \hfil \today\quad\militarytime}
        \let\@evenhead\@oddhead
        \let\label=\draftlabel
        \let\marginnote=\draftmarginnote
        \def\ps@empty{\let\@mkboth\@gobbletwo
        \def\@oddfoot{\hfil \smash{\Large\sl DRAFT} \hfil}
        \let\@evenfoot\@oddhead}
        \def\@eqnnum{(\theequation)\rlap{\kern\marginparsep\tt\@eqnlabel}%
        \global\let\@eqnlabel\@vacuum}  }
\begin{document}

\begin{titlepage}
{\hbox to\hsize{June 2002 \hfill {AEI--2002--041}}} {\hbox to\hsize{${~}$\hfill
{Bicocca-FT-02-9}}} {\hbox to\hsize{${~}$ \hfill {CERN--TH/2002-120}}} {\hbox
to\hsize{${~}$ \hfill {IFUM--717--FT}}} {\hbox to\hsize{${~}$ \hfill
{LAPTH-916/02}}}
\begin{center}
\vglue .06in {\Large\bf Non-protected operators in ${\cal N}=4$ SYM and
multiparticle states of AdS$_5$ SUGRA}
%\\
%\today
\\[.3in]
{\large {$\rm{G. Arutyunov}^{(a)}$}}, {\large{$\rm{S. ~Penati}^{(b)}$, $\rm{A.
C. ~Petkou^{(c)}}$, $\rm{A. ~Santambrogio^{(d)}}$, $\rm{E.
~Sokatchev^{(c,e)}}$}}
\\[.3in]
\small

$^{\rm(a)}$ {\it Max-Planck-Institut f\"ur Gravitationsphysik,
Albert-Einstein-Institut, \\
Am M\"uhlenberg 1, D-14476 Golm, Germany}
\\
[.03in]
$^{\rm(b)}${\it Dipartimento di Fisica, Universit\`a degli studi di
Milano-Bicocca and INFN, Sezione di Milano, piazza della Scienza 3,
I-20126 Milano, Italy}
\\
[.03in] $^{\rm(c)}${\it CERN, Theory Division, 1211 Geneva 23, Switzerland}
\\
[.03in] $^{\rm(d)}${\it Dipartimento di Fisica, Universit\`a degli studi di
Milano and INFN, Sezione di Milano, via Celoria 16, I-20133 Milano, Italy}
\\
[.03in] $^{\rm(e)}${\it Laboratoire d'Annecy-le-Vieux de Physique Th\'{e}orique
LAPTH, B.P. 110, F-74941 Annecy-le-Vieux et l'Universit\'{e} de Savoie}
\\[.3in]
\normalsize

{\bf ABSTRACT}\\[.0015in]
\end{center}

We study a class of non-protected local composite operators which occur in the
R symmetry singlet channel of the OPE of two stress-tensor multiplets in ${\cal
N}=4$ SYM. At tree level these are quadrilinear scalar dimension four
operators, two single-traces and two double-traces. In the presence of
interaction, due to a non-trivial mixing under renormalization, they split into
linear combinations of conformally covariant operators. We resolve the mixing
by computing the one-loop two-point functions of all the operators in an ${\cal
N}=1$ setup, then diagonalizing the anomalous dimension matrix and identifying
the quasiprimary operators. We find one operator whose anomalous dimension is
negative and suppressed by a factor of $1/N^2$ with respect to the anomalous
dimensions of the Konishi-like operators. We reveal the mechanism responsible
for this suppression and argue that it works at every order in perturbation
theory. In the context of the AdS/CFT correspondence such an operator should be
dual to a multiparticle supergravity state whose energy is less than the sum of
the corresponding individual single-particle states.

\vskip 3pt
${~~~}$ \newline
%\\[.01in]

\end{titlepage}

\section{Introduction}

One of the fundamental ideas driving much of the progress in theoretical high
energy physics over the past decades is the perceived deep relation between
gauge fields and strings. Perhaps the most concrete realization of this idea is
the AdS/CFT correspondence \cite{Maldacena:1997re,Gubser:1998bc,Witten:1998qj}
that conjectures a duality between, for example, type IIB string theory on
AdS$_5\times$S$^5$ and ${\cal N}=4$ $SU(N)$ super Yang--Mills in four
dimensions. A lot of evidence for this conjecture  has been accumulated during
the past few years (for an exhaustive list of references,
see e.g. \cite{Aharony:1999ti,D'Hoker:2002aw}).

A large part of the tests for the validity of the AdS/CFT correspondence are
based on the comparison of perturbative (weak coupling) calculations in ${\cal
N}=4$ SYM to results obtained from AdS$_5 \times S^5$ supergravity, the latter
supposedly describing the theory at strong t'Hooft coupling $\lambda
=g_{YM}^2N/(4\pi^2)$.\footnote{Strictly speaking, such calculations test the
weak version of the AdS/CFT correspondence that conjectures a duality between
the large $N$, large-t'Hooft coupling limit of ${\cal N}=4$ SYM and the SUGRA
limit of IIB string theory on AdS$_5\times$S$^5$. Recently there have been very
interesting attempts to include massive strings modes of IIB theory in such
tests \cite{Berenstein:2002jq,Gubser:2002tv}.}

It is natural to start by studying quantities having the simplest possible
behavior at weak and strong coupling. These include, for instance, two- and
three-point correlation functions of chiral primary operators (CPOs), $n$-point
correlators of the ``extremal" and ``near-extremal" type, as well as conformal
and R-current anomalies. These quantities are essentially related to BPS states
(short, or protected operators) in either the gauge or the supergravity/string
sectors, and therefore they render only a part of the correspondence between
gauge theories and strings.

%\footnote{A non-trivial test of the AdS/CFT
%correspondence has been presented e.g. in \cite{Eden:2000bk} but the
%information contained in it has not been decoded yet.}

If we would like to go further and use the AdS/CFT duality to better understand
the relation between gauge and string theories, we should consider
non-protected quantities, non-BPS operators in particular. In string theory
there is a clear distinction between supergravity (light) and string (heavy)
modes. From this point of view non-BPS operators are naturally interpreted as
falling into two different classes: operators from the first class are dual to
``gravity multiparticle states"\footnote{Multiparticle supergravity states
constitute a wide class including, e.g., states dual to 1/4 BPS operators of
the gauge theory.}, while operators from the second class are dual to ``string
single/multiparticle states". To which class a concrete operator belongs can be
decided by looking at the behavior of its scaling dimension at strong coupling
since the latter is related by duality to the mass of the corresponding
particle. This is a nice picture but hardly operational because of our present
inability to compute scaling dimensions of non-protected operators {\it at
strong coupling} by QFT methods. It is therefore of certain interest to see if
one can classify non-BPS operators by their different {\it weak-coupling}
behaviour and make contact with their string theory interpretation.

In the present work we focus on a small but controllable part of the
non-protected sector of operators of ${\cal N}=4$ SYM. These are superconformal
primaries that appear in the OPE of two CPOs with dimension two in the
$[0,2,0]$ representation of the $SU(4)\sim SO(6)$ R symmetry group (the lowest
components of the stress-tensor multiplet). It has been shown
\cite{Arutyunov:2001qw,Heslop:2001dr} that non-protected superconformal
primaries appear {\it only} in the singlet channel of that OPE. In free field
theory, this infinite set of non-protected operators naturally breaks into two
classes. One of them, which we call the $K$-class, consists of operators which
are bilinear single-traces in the elementary scalar fields while the other, the
$O$-class, consists of operators that are quadrilinear double-traces. An
important property of the operators in the $K$-class is that their canonical
dimensions saturate the unitarity bound
\cite{Ferrara:1998ej,Andrianopoli:1998jh,Dobrev:qv}, i.e., they are twist-two
operators, and in free field theory they are conserved tensors (except the
scalar $K$-class operators) \cite{Todorov:rf,Fradkin:is}. The operators in the
$O$-class, instead, do not obey any conservation condition in free field theory
and their canonical dimensions are above the unitarity bound, i.e., they are
higher-twist operators.

It is one of the important features of AdS/CFT that the operators in the above
two classes are clearly distinguished by the large $N$ strong coupling dynamics
of the ${\cal N}=4$ SYM. Namely, the operators in the $K$-class should acquire
large anomalous dimensions in the $\lambda\rightarrow\infty$ limit and, as a
consequence, they decouple in the supergravity regime. This is thought to be a
manifestation of the fact that the $K$-class operators are dual to massive
string modes \cite{Gubser:1998bc}. On the other hand, in what appears to be one
of the more intriguing  non-trivial AdS/CFT results, the contributions of {\it
some of } the operators in the $O$-class survive in the supergravity induced
four-point function of stress-tensor multiplets, while at the same time these
operators receive non-trivial corrections to their canonical dimensions both
perturbatively and in the supergravity regime
\cite{Bianchi:1999ge,D'Hoker:1999jp,Arutyunov:2000py,Eden:2000mv,Bianchi:2000hn,Arutyunov:2000ku,Arutyunov:2000im,Eden:2000bk}.
The exact reason which stays behind this behavior remains unclear. It is
commonly stated that the operators in the $O$-class which do not drop out in
the strong coupling limit are associated with multiparticle supergravity
states. No matter how appealing such a point of view might be, it is at the
same time rather puzzling since it is not immediately clear how non-trivial
tree level corrections to the energy of multiparticle states could arise in a
theory like the AdS$_5 \times S^5$ supergravity where the energy of all
single-particle states is protected.

In the present work, we intend to shed more light on the above question by
studying the simplest scalar operators in the $O$-class, i.e. (free field)
dimension 4 scalar operators occurring in the singlet $R$--symmetry channel in
the OPE of two stress-tensor multiplets. Our interest in looking at this
particular example is motivated by the knowledge of the OPE structure in the
supergravity approximation. As was found in Ref. \cite{Arutyunov:2000ku}, the
$SU(4)$-singlet channel of the corresponding OPE contains  a scalar operator of
dimension $\Delta=4+\gamma$, where the strong-coupling anomalous dimension is
given by $\gamma=-16/N^2$. Thus, this operator, which survives in the
supergravity approximation, is a natural candidate to be identified with the
gravity two-particle bound state. It is therefore interesting to see if one can
find the weak coupling counterpart of this operator by studying the
corresponding OPE in perturbation theory.

In the free field limit, the dimension four scalar operators we are interested
in are suitable linear combinations of the four possible independent
quadrilinear structures, two single traces and two double traces, which can be
chosen to be orthonormal.

Switching on the interaction, we study their quantum properties by evaluating
their one-loop two-point functions. UV infinities which arise at this order
require a non-trivial renormalization of the four operators and the consequent
appearance of anomalous dimensions. The evaluation of the anomalous dimensions
is complicated by a non-trivial mixing of the operators under renormalization.
The complete anomalous dimensions matrix has to be determined and from its
diagonalization we can read off the anomalous dimensions of the four ``pure CFT
states''. The latter correspond to the correct {\it quasiprimary} operators in
the $O$-class. Our main result is that {\it one} of the above four quasiprimary
operators acquires an anomalous dimension which is suppressed by a factor of
$1/N^2$ with respect to the anomalous dimensions of {\it all} the operators in
the $K$-class, as well as with respect to the remaining three quasiprimary
operators in the $O$-class. The mechanism behind this suppression is tied to
the fact that, at large $N$, the dominant sector in the two-point function
calculation for the operator in question is given by planar disconnected
Feynman graphs. The latter factorize into two-point functions of CPOs, i.e., of
operators {\it protected from renormalization}. Consequently, the leading
contributions vanish and one is left with a mixture of planar and non-planar
graphs, which in our particular case are of the same order in $1/N$. This
``partial non-renormalization" argument applies to all orders in perturbation
theory. Remarkably, the anomalous dimension of the above quasiprimary operator
is negative (at one loop, at least) and can be interpreted as a binding energy.
This is in accordance with the interpretation of this operator as corresponding
to a two-particle bound state with energy smaller than the sum of the energies
of the respective single-particle states.  We point out that the strong
coupling anomalous dimension of the surviving $O$-class operator(s) is non-zero
and negative which can also be interpreted as a binding energy of the
corresponding two-particle supergravity state. Our results indicate that both
planar and non-planar graphs are responsible for the non-trivial anomalous
dimensions of supergravity states. This raises an interesting question
regarding the nature of the binding energies of multiparticle supergravity
states. Similar questions have been recently discussed in
\cite{Aharony:2001pa}.

The organization of the paper is as follows. In Section 2 we discuss the free
field theory OPE of two lowest weight CPOs. In Section 3 we present the general
formalism for calculating perturbative anomalous dimensions of composite
operators in CFT and for dealing with the problem of operator mixing and
splitting. In Section 4 we recall the ${\cal N}=1$ superspace approach
\cite{Penati:1999ba,Penati:2000zv} that we use to calculate the one-loop
two-point functions of the four free-field operators and fix our notations and
conventions. This formalism is then used in Section 5 to compute the anomalous
dimension matrix. The diagonalization procedure is then performed to obtain the
corresponding one-loop quasiprimary operators. We test our results for
consistency with certain constraints obtained from the independent OPE analysis
of one- and two-loop four-point calculations. In Section 6 we discuss our
results in connection with the AdS/CFT duality. In Appendix A we provide some
relevant information about the perturbative four-point functions of the lowest
weight CPOs and the corresponding OPE. Appendix B contains some technical
details of the calculations.

\section{Non-protected operators in  ${\cal N}=4$ SYM}

There is by now an exhaustive list of gauge invariant protected operators in
${\cal N}=4$ SYM, both of the BPS type \cite{Andrianopoli:1999vr} and of the
so-called ``semishort" type (see, e.g., \cite{Ferrara:2001uj}). Non-protected
gauge invariant operators are much less well understood, even if they form
perhaps the most important part of the spectrum. In this work we focus on a set
of non-protected operators of ${\cal N}=4$ SYM that is controllable and
relatively well known. These are the superconformal primaries that appear in
the OPE of two CPOs with lowest dimension. The latter are the lowest components
of the supercurrent multiplet and are formed by single trace bilinears of the
six real scalars $\phi^L(x)$, $L=1,2,\ldots,6$ of ${\cal N} =4$ SYM
transforming under the ${\bf 20}\equiv [0,2,0]$ irrep of $SU(4)$. Explicitly,
up to a normalization constant, they are defined as \be \label{O20} O_{\bf
20}^{LM}(x) \sim
  \Tr[\phi^L(x)\phi^M(x)]
-\frac{1}{6}\delta^{LM}\Tr[\phi^N(x)\phi^N(x)] \ee where the trace is over the
$SU(N)$ indices.  The OPE of two such operators involves all operators in the
decomposition \bea \label{decomp} [0,2,0]\times [0,2,0] &=& [0,0,0] +[0,2,0] +
[0,4,0] + [2,0,2]
\nonumber \\
 & & +[1,0,1] + [1,2,1]
\eea The irreps in the first line are realized as even spin symmetric traceless
tensors while those in the second line as odd spin ones. A general
non-renormalization theorem then shows that among the superconformal primaries
that may appear in the OPE of the two CPOs (\ref{O20}), {\it only} the ones in
the singlet $[0,0,0]$ channel are non-protected by superconformal invariance
\cite{Arutyunov:2001qw,Heslop:2001dr}. For the latter operators the only
constraint imposed by superconformal invariance is the unitarity condition
\cite{Ferrara:1998ej,Dobrev:qv} \be \label{unit} \Delta \geq 2 +s \ee where
$\Delta$ is the scaling dimension of the corresponding operator while
$s=2n\,,\,n=0,1,2,..$ is the spin.

An explicit expression for the singlet channel of the free OPE can then be
written as follows: \bea O^{LM}_{\bf 20}(x_1)O^{LM}_{\bf 20}(x_2) &=&
\frac{1}{x_{12}^4} + a\frac{1}{x_{12}^2} \d_{LM}:\!{\rm
Tr}\left[\phi^L(x_1)\phi^M(x_2)\right]\!:
~+~ b :\!O^{LM}_{\bf 20}(x_1)O^{LM}_{\bf 20}(x_2)\!:\nonumber \\
\label{frOPE}&=&\frac{1}{x_{12}^4} +
\frac{1}{x_{12}^2}C_K(x_{12},\partial_2)*[K](x_2)
+C_O(x_{12},\partial_2)*[O](x_2) \eea where $a$ and $b$ are constants dependent
on the normalization of the operators. $[K](x)$ and $[O](x)$ denote the $K$-
and $O$-class operators, respectively and  $C(x_{12},\partial_2)$ denote the
corresponding OPE coefficients \cite{Arutyunov:2000ku}. The important
assumption behind a conformal OPE such as (\ref{frOPE}) is that the operators
appearing in the r.h.s. are ``pure CFT states'' or else {\it quasiprimary
operators}. This means that under conformal transformations they behave in a
well-defined way determined uniquely by their spin and scaling dimension. In
practice, quasiprimary operators form an orthogonal basis of the operator
algebra such that all two-point functions between different operators vanish
identically. This latter property, together with the explicit expressions for
all the OPE coefficients $C(x_{12},\partial_2)$ \cite{Hoffmann:2000mx} allows
one to study conformally invariant four-point functions in terms of the OPE.

It should be pointed out, however, that sometimes the conformal labels (spin
and dimension) are not sufficient to distinguish all the operators of a given
type. For instance, later on in this paper we will have to deal with the
degeneracy of the free quadrilinear scalar operators of dimension four: We will
find four such structures, all with the same conformal labels. It is generally
believed that the conformal interactions lift such degeneracies by creating
{\it anomalous dimensions}. Indeed, this is what will happen in our case.

{}From (\ref{frOPE}) we see that the free-field realization of the $K$-class
operators is given in terms of single-trace bilinears while that of the
$O$-class operators in terms of double-trace quadrilinears in the $\phi$'s.
Consequently, the canonical dimensions of the operators in the $K$-class are of
the form \be \label{twist2} \D_K = 2+s\,,\,\,\quad s=0,2,4,... \ee while those
in the $O$-class are of the form \be \label{twist4} \D_O = 4+s\,,\,\,\,\quad
s=0,2,4,... \ee In other words, the operators in the $K$-class are of twist two
while those in the $O$-class are of twist four. Furthermore, since the
canonical dimensions of the free $K$-class operators saturate the unitarity
bound (\ref{unit}), these operators  form an infinite set of free short
supermultiplets containing conserved currents. However, the corresponding
supermultiplets become long in the interacting theory since there is no
mechanism to protect their dimensions from  radiative corrections
\cite{Ferrara:1998ej}. The operators in the $O$-class do not saturate the
unitarity bound (\ref{unit}) and they also acquire in general anomalous
dimensions.

Now, if the conformal OPE (\ref{frOPE}) is valid as an operator statement one
should be able to define the operators in the $K$- and $O$-classes, both in
perturbation theory as well as non-perturbatively, and study their properties.
This is clearly a formidable task. For the operators in the $K$-class this
would be very interesting in view of the recent conjecture regarding the
behavior of the anomalous dimension of operators with twist two with large
canonical dimension and large spin \cite{Gubser:2002tv}.\footnote{The one-loop
anomalous dimensions $\eta_K=\D_K-2-s$ of the operators in the $K$-class have
been calculated in \cite{Dolan:2001tt} by applying conformal OPE techniques to
 the one-loop four-point function of CPOs:
$$ \eta_{K} =\frac{\lambda}{2\pi^2}\sum_{k=0}^{s+2}\frac{1}{k}\sim
\frac{\lambda}{2\pi^2}\ln s \,\,\,\,\,\,\, \mbox{for $s\rightarrow\infty$} $$
Using the results of \cite{Arutyunov:2001mh} one could extend the above
analysis to two loops and test the conjecture of \cite{Gubser:2002tv} about the
absence of $\ln^k s$ terms in the $k$-loop anomalous dimension of the $K$-class
operators.} The study of higher-twist composite operators is equally
interesting and its the purpose of the present work to initiate the
investigations in this direction.

\section{Anomalous dimensions and operator mixing in CFT}

The study of composite operators in a renormalizable quantum field theory is
among the most important and complicated subjects \cite{Osborn:gm}. Even if
such studies simplify considerably in CFTs due to the absence of beta
functions, they remain quite elaborate due to operator mixing under
renormalization. In this Section we present the general framework for the study
of composite operators in CFT in the presence of operator mixing, setting up
the explicit calculations of the following Sections.

\subsection{Anomalous dimensions of composite operators in CFT}

Given  a renormalizable QFT, we consider a renormalized composite operator
${\cal O}(x)$ (scalars only, for simplicity) constructed from the elementary
fields of the theory. The fact that the operator is renormalized means that the
insertions of ${\cal O}(x)$ into renormalized correlation functions are finite.
In particular, the $n$-point functions of ${\cal O}(x)$ are finite. If the
theory is at its fixed point (CFT, the beta functions vanish) the renormalized
$n$-point functions satisfy Ward identities which become the configuration
space analog of the Callan-Symanzik RG equations at the fixed point
\cite{Zamolodchikov:ai} \be \label{WI2} \left[\sum_{k=1}^n x^{\m}_k
\frac{\pa}{\pa x^\m_{k}}
  +n\Delta\right] \langle {\cal O}(x_1)...{\cal O}(x_k)...\rangle_R = 0\,.
\ee
The parameter $\D$ is the scaling dimension of the operator
${\cal O}(x)$ and determines its transformation properties under
scale
  transformations.
In particular, for the two-point function we have
\be
\label{2ptWI}
\left[x^{\m} \frac{\pa}{\pa x^\m}
  +2\Delta\right] \langle {\cal O}(x){\cal O}(0)\rangle_R = 0
\ee In a free field theory, the dimension $\Delta$ appearing in (\ref{2ptWI})
coincides with the {\it canonical} dimension $\Delta_0$ of the operator which
can be inferred from the canonical dimensions of the elementary fields entering
the definition of ${\cal O}$. Translation invariance together with the Ward
identity (\ref{2ptWI}) then constrain the two-point functions to be power-like.
After turning on the interaction, the perturbative evaluation of the two-point
function leads in general to divergences which emerge from the explicit
calculation of Feynman integrals. Since the subtraction of these divergences
introduces a non-trivial dependence on the UV scale, the two-point function
looses its power-like behavior and does not satisfy (\ref{2ptWI}) anymore.
However, order by order in the perturbative expansion we can recover the
conformal Ward identity by modifying the dimension as $\Delta= \Delta_0 +
\gamma$. The deviation $\gamma$ from the canonical dimension is the {\it
anomalous dimension} which is given as a power series in the perturbation
parameter.

To be more explicit and to show how the general calculation works, we consider
the bare (non-renormalized) field ${\cal O}^{(0)}(x)$ and evaluate $\langle
{\cal O}^{(0)}(x){\cal O}^{(0)}(0)\rangle $ in perturbation theory. At tree
level we recover the power-like behavior\footnote{Notice that it is
  possible that renormalized conformal two-point functions may contain
  divergences even at tree level. These divergences are due to
  {\it ultralocal} short distance singularities in the two-point
  function (see, e.g., \cite{Osborn:cr}). They are
  insensitive to the perturbative expansion \cite{Petkou:1999fv} and
  give rise to
  {\it conformal anomalies} which should not be confused with the {\it
    anomalous} dimensions of the operators.}. The assumption that the
theory is a {\it non-trivial} renormalized CFT requires the existence of a
dimensionless renormalized coupling $\lambda$. Using dimensional regularization
($d = 4 - 2\e$) to regularize the integrals, we can present the UV divergences
as simple poles in $\e$, e.g., at one loop \be \label{1loop} \langle {\cal
O}^{(0)}(x){\cal O}^{(0)}(0)\rangle|_{0+1} = \frac{1}{(x^2)^{\Delta_0}}[1+\l
\frac{a}{\epsilon} + O(\e^0)] \ee where $a$ is a constant independent of $\l$
and $\e$. We also work henceforth with normalized conformal operators. To
cancel the divergence in (\ref{1loop}) we perform the usual multiplicative
renormalization of the composite operator by introducing a divergent
renormalization constant $Z(\l,\epsilon)$ \be \label{renorm} {\cal O}(x) \equiv
Z(\l,\epsilon){\cal O}^{(0)}(x) \ee which is determined by requiring $\langle
{\cal O}(x){\cal O}(0)\rangle $ to be finite at this order. Using the minimal
subtraction scheme, this immediately gives \be Z(\l,\epsilon) = 1 - \frac{\l
a}{2} \, \frac{1}{\e} \label{Zfunct} \ee Since $Z(\l,\e)$ depends on the UV
scale, the one-loop renormalized two-point function does not satisfy the
conformal Ward identity (\ref{2ptWI}). To remedy this we postulate that the
conformal Ward identity (\ref{2ptWI}) is satisfied order by order in
perturbation theory and at the same time the canonical dimension of the
operator $O(x)$ is modified as \be \label{Delta1} \Delta^{(1)} =\Delta_0
+\gamma^{(1)} \ee Then, $\g^{(1)}$ is determined by requiring \be \label{2pt1}
\left[ \m\frac{\pa}{\pa \m}
  +2\Delta^{(1)}\right] \langle {\cal O}(x){\cal O}(0)
\rangle|_{0+1} = O(\l^2) \ee where $\mu$ is the mass scale of dimensional
regularization. To leading first order in $\l$, we obtain from (\ref{2pt1}) \be
\label{anomd} \gamma^{(1)} = -\mu\frac{\partial}{\partial\mu}\ln Z \ee The
$\mu$-dependence of $Z$ comes from dimensional transmutation since in $(4-2\e)$
dimensions the physical coupling $\l_*$ becomes dimensionful and we need to
define a dimensionless coupling as $\l = \m^{2\e}\l_*$. As a consequence, we
find \be \g^{(1)} = \l a \ee This is the first-order version of the more
general, well--known result that the anomalous dimension is read off from the
coefficient of the simple pole in the renormalization constant $Z$. The
solution of (\ref{2pt1}) at this order in perturbation theory is \be \label{OR}
\langle {\cal O}(x){\cal O}(0)\rangle_{0+1} = \frac{1}{x^{2\Delta_0}}[1+ C \l
-\gamma^{(1)}\ln(x^2\mu^2)] \ee where the constant $C$ encodes possible finite
renormalization (finite counterterms) of the two-point function. This relation
is an explicit proof of the statement that the modification of the canonical
dimensions of quantum operators is due to renormalization and is signaled by
logarithmic terms appearing in the two--point functions.

\subsection{Operator mixing in CFT}

\vskip 15pt The general discussion presented above cannot be immediately
applied when the free theory contains more than one linearly independent
composite operators having the same canonical dimensions. In fact, in such a
case one can start in the free field theory with a set of mutually orthogonal
operators which correspond to a basis of free quasiprimary operators. However,
in general nothing prevents them from mixing at the quantum level. Therefore,
in perturbation theory one has to orthonormalize the basis order by order, as
we are going to describe in detail \footnote{In the context of ${\cal N}=4$
SYM, the mixing of composite operators has been discussed in a number of recent
papers
\cite{D'Hoker:1999jp,Bianchi:1999ge,Arutyunov:2000ku,Bianchi:2001cm,Ryzhov:2001bp}.}.

Suppose that we have a set of scalar operators ${\cal O}_i$, $i=1,\ldots,p$ of
equal free dimension $\Delta_0$ and with identical quantum numbers. By suitable
rescalings of the operators we can choose this set to be orthonormal \be
\label{mix1} \langle {\cal O}^{(0)}_i(x){\cal O}^{(0)}_j(0)\rangle_0 =
\frac{\delta_{ij}}{x^{2\Delta_0}} \ee Note that this choice of basis is not
unique as we can always make an orthogonal transformation ${\cal O}'_i =
o_{ij}{\cal O}_j$, $o^To = \mathbb{I}$ which preserves (\ref{mix1}). Now, let
us switch on the interaction and compute the two-point functions at order $\l$.
In general, at one-loop we expect to find something like \be \label{mix1'}
\langle {\cal O}^{(0)}_i(x){\cal O}^{(0)}_j(0)\rangle_{0+1} =
\frac{1}{x^{2\Delta_0}} [\delta_{ij} + \l \rho_{ij}
  + \l \frac{\omega_{ij}}{\e} + O(\e)]
\ee
where $\rho=\rho^T$ and $\omega = \omega^T$ are constant symmetric
matrices\footnote{Here we make the discussion more general by
  considering also finite
contributions $\rho$ which correspond to finite renormalizations of the
two--point function.}. In general they are not diagonal, as a result of the
fact that at one loop the operators are not orthogonal anymore and develop
mixed, possibly divergent, two--point functions.

To cancel the $1/\e$ poles we introduce a renormalization matrix $Z_{ij}$ as
\be {\cal O}_i \equiv Z_{ij} O^{(0)}_j\,,\,\,\,\,\, Z_{ij} = \d_{ij} -
\frac{1}{\e}\frac{\l \omega_{ij}}{2} \ee which is determined by requiring
$\langle {\cal O}_i{\cal O}_j \rangle_{0+1}$ to be finite. If we formally
follow the same reasoning which in the case of a single operator brought us to
(\ref{OR}), we eventually arrive at the set of renormalized one--loop
two--point functions \be \label{mix2} \langle {\cal O}_i(x){\cal
O}_j(0)\rangle_{0+1} = \frac{1}{x^{2\Delta_0}} [\delta_{ij} + \l(\rho_{ij}
  - \omega_{ij} \ln{ (x^2 \m^2)})] + O(\l^2)
\ee The above result shows that in general the one-loop renormalized operators
in (\ref{mix2}) fail to be quasiprimary, since the two-point functions of the
latter should be diagonal, \be \label{mix3}
  \langle \Tilde{\cal O}_i(x) \Tilde{\cal O}_j(0)\rangle_{0+1} =
  \frac{\delta_{ij}}{x^{2\Delta_0}}
  [1 +\l \rho_i- \l \g_i \ln{ (x^2\m^2)}]  + O(\lambda^2)
\ee
(no summation on repeated indices is assumed).
Therefore, in order to find the explicit realization of the one-loop
quasiprimary operators we need to bring eq. (\ref{mix2}) in the form
(\ref{mix3}). This can be done by performing the linear transformation
\be
\Tilde{\cal O} = (L_0 + \l L_1) {\cal O}
\label{mix4}
\ee
which implies
\bea
\label{mix15}
&&\langle \Tilde{\cal O}(x)\Tilde{\cal O}(0)\rangle =
(L_0 + \l L_1) \langle {\cal O}{\cal O}\rangle
(L_0 + \l L_1)^T  \\
&&=\frac{1}{(x^2)^{\Delta_0}}\left[ L_0 L_0^T +
\l (L_1 L_0^T + L_0 L_1^T +
 L_0\rho L_0^T - L_0\omega L_0^T \ln{ (x^2 \m^2)}) \right] +
O(\lambda^2)\nonumber
\eea
Comparing (\ref{mix3}) with (\ref{mix15}) term by term, we obtain the
following three equations:
\begin{eqnarray}
  && L_0 L_0^T = \mathbb{I} \label{mix5}\\
  &&L_0\omega L_0^T = \G  \label{mix6}\\
  && L_1 L_0^T + L_0 L_1^T +
 L_0\rho L_0^T = P \label{mix7}
\end{eqnarray}
where $\G$ denotes a diagonal matrix with eigenvalues $\g_i$ and $P$
a diagonal matrix with eigenvalues $\rho_i$. The corresponding
eigenvectors $\Tilde{\cal O}_i$ are the one-loop quasiprimary
operators having one-loop anomalous dimensions $\g_i$.

{}From (\ref{mix5}) we see that the matrix $L_0$ must be orthogonal and
from (\ref{mix6}) we see that it must diagonalize the matrix
$\omega$. If all the
eigenvalues $\g_i$ are different, $\g_i\neq \g_j$ (i.e., all the
quasiprimary operators
acquire different anomalous dimensions), then the solution of eq.
(\ref{mix6}) is {\em unique}. In this case, the one-loop
renormalization and diagonalization of the anomalous dimension matrix
{\it uniquely fixes} the form of the tree level quasiprimary operators.
Finally, eq. (\ref{mix7}) involves the finite counterterms
$\rho_i$. These can be fixed by choosing a suitable normalization
for the one-loop renormalized correlators (\ref{mix3}). Choosing for
example $\rho_i=0$ ($P=0$) and setting $L_1 = \ell L_0$, eq. (\ref{mix7})
can be rewritten as
\be
\label{mix8}
  \ell + \ell^T = - L_0\rho L_0^T
\ee
We notice that this equation leaves some freedom that may affect
the explicit expression of the one-loop quasiprimary operators to
$O(\l)$. To fix this freedom one would need to discuss the two-point
functions (\ref{mix2}) at order $O(\l^2)$, but this is outside the
scope of the present work. A similar pattern is expected to be reproduced
order by order.

We conclude that the diagonalization of the set of operators at one loop
requires a particular orthogonal transformation $L_0$ already at tree level.
This means that the choice of the free basis is not arbitrary, but is
completely determined by the one-loop $\ln{(x^2 \m^2)}$ terms in the full
matrix of two-point functions. Therefore, when discussing an operator at the
free level with an eye to its quantum properties, we need to resolve the mixing
as explained above.

\subsection{Operator splitting in CFT}

Let us now discuss what are the consequences of the above mixing phenomenon for
the OPE and the four-point functions. Considering for clarity only scalar
fields, the generic form of a free OPE is \cite{Arutyunov:2000ku} \be
\label{OPE1} A(x_1)A(x_2) = \frac{1}{x_{12}^{2\Delta_A}} + g_{O} \,
\frac{1}{x_{12}^{2(\Delta_A
    -\frac{1}{2}\Delta_O)}}[O(x_2)+\cdots ]
\ee
where the dots represent derivatives of $O$ and
$\Delta_A\,,\,\Delta_O$ are the canonical dimensions of $A$ and
$O$ respectively.
The parameter $g_O$ is connected to the three-point function of $O$ as
\beq
\label{O2}
\langle A(x_1)A(x_2)O(x_3)\rangle =
\frac{g_O}{x_{12}^{2(\Delta_A-
    \frac{1}{2}\Delta_O)}(x_{13}^2x_{23}^2)^{\frac{1}{2}\Delta_O}}
\ee The important point in free field theory is that given the explicit
expression for $A$ in terms of the elementary fields, one uniquely obtains the
expressions of all the fields on the r.h.s. of the OPE (\ref{OPE1}) using for
example a Taylor expansion of the l.h.s. \cite{Lang:1992pp,Arutyunov:2000ku}.
In fact, to any given dimension and spin corresponds {\it one} explicit
expression of elementary fields, such as $O$ in (\ref{OPE1}). However,
complications arise when there exists at the free theory level more than one
linearly independent expressions in terms of  the elementary fields
corresponding to conformal operators of a given dimension and spin. In this
case the fields appearing on the r.h.s. of the OPE (\ref{OPE1}), such as $O$,
represent a particular linear combination of the different, linearly
independent expressions. This phenomenon is termed the {\it free-field theory
degeneracy} of the conformal OPE.

In the presence of such a degeneracy one is forced to study the mixing under
renormalization of all the linearly independent expressions, following the
discussion presented before. In this way, at one-loop, one ends up with a set
of quasiprimary operators which should be the correct CFT states appearing in
the OPE (\ref{OPE1}). Finally, inverting relations such as (\ref{mix4}) one can
express $O$ in terms of an orthonormal basis of one-loop quasiprimary operators
$\S_i$ \be \label{Ofree} O(x) = \sum_{i=1}^{p}a_i\,\S_i(x) \ee where $p$ is the
number of the linearly independent free field expressions. At one loop the
coefficients $a_i$ in (\ref{Ofree}) are just numbers (they do not depend on the
coupling) that are uniquely determined by the diagonalization procedure
explained before. The relation (\ref{Ofree}) is the {\it operator splitting}
discussed in \cite{Arutyunov:2000ku,Arutyunov:2000im} in the case of the stress
tensor of ${\cal N}=4$ SYM.

Once the mixing of the various conformal operators described in the previous
subsection has been resolved, we can make contact with the independent
four-point function approach to the OPE. In it one tries to extract information
about the spectrum and the anomalous dimensions of the operators involved in
the OPE by making a conformal partial wave expansion of the four-point
amplitude of, e.g., four operators $O_{\bf 20}$. Matching the terms $\ln v$,
$\ln^2 v$, etc. in this expansion ($ v= (x_{12}^2x_{34}^2)/(x_{14}^2x_{23}^2)$
is one of the conformal cross-ratios), one can find consistency conditions on
the mixing coefficients $a_i$ in (\ref{Ofree}) and on the anomalous dimensions.
The above four-point amplitude has been computed up to two loops
\cite{Eden:2000mv}, which allows one to find three such conditions. This
procedure has already been discussed in \cite{Arutyunov:2001mh,Bianchi:2001cm}, but the full
details needed for its application are worked out in Appendix A. Assuming that
the quasiprimary operators $\S_i$ have one-loop anomalous dimensions $\g_i$ and
are normalized, we can write down the following consistency relations \bea
\label{cons1}
\sum_{i=1}^p a^2_i &=& 1\\
\label{cons2}
\sum_{i=1}^p a^2_i\g_i &=& -4\frac{\lambda}{N^2} \\
\label{cons3} \sum_{i=1}^p a^2_i\g_i^2 &=& 18\frac{\lambda^2}{N^2} \eea The
first of these conditions can be viewed merely as a conventional normalization
of the operator $O$ in (\ref{Ofree}). The second is derived from the one-loop
and the third from the two-loop four-point amplitude (at order $1/N^2$).
Although these equations are not sufficient to determine all the eight
parameters $a_i,\gamma_i$ (in our case $p=4$, see the next section), they
nevertheless put some non-trivial constraints on the parameters. For instance,
since the right-hand side of eq. (\ref{cons2}) is negative and is of order
$1/N^2$, we expect to find in our calculation that: i) at least one anomalous
dimension is negative; ii) all the quasiprimary operators corresponding to
finite (in the large $N$ limit) anomalous dimensions enter in the linear
combination for $O_1$ with coefficients of order at least $1/N$. In addition,
conditions (\ref{cons1})--(\ref{cons3}) provide a very non-trivial cross-check
both on our two-point one-loop calculations and on the existing perturbative
four-point ones. In Section 5 we show that our results do indeed verify these
conditions.

\section{${\cal N}=1$ superspace approach}

We are interested in the calculation of the anomalous dimensions of
scalar composite operators of ${\cal N}=4$ SYM theory
which have canonical dimension $4$
and belong to the $[0,0,0]$ representation of the R--symmetry group.
In particular,
we can construct four possible dimension 4 quadrilinear operators
in terms of the six elementary scalar fields of the theory
\bea
\label{mix9}
{\cal A}_{ 1} &=& {\rm Tr}(\phi^L\phi^M){\rm Tr}(\phi^L\phi^M)  \nonumber\\
{\cal A}_{ 2} &=& \left[{\rm Tr}(\phi^L\phi^L)\right]^2  \nonumber\\
{\cal A}_{ 3} &=& {\rm Tr}(\phi^L\phi^M\phi^L\phi^M)  \nonumber\\
{\cal A}_{ 4} &=& {\rm Tr}(\phi^L\phi^L\phi^M\phi^M),
\eea
As explained in Subsection 3.3, the free OPE (\ref{frOPE}) contains
one particular combination - named ${\cal O}_1$ - of
the above operators. This is obtained by explicit calculations  as
\cite{Arutyunov:2000im}
\be
\label{OPEO1}
O^{LM}_{\bf 20}(x_1)O^{LM}_{\bf 20}(x_2) \sim \frac{1}{x_{12}^4}
+ [{\cal O}_1(x_2)+(\partial {\cal O}_1(x_2))] +\cdots
\ee
with
\be
\label{O1def}
  {\cal O}_{ 1} \sim {\cal A}_{ 1} - \frac{1}{6}{\cal A}_{ 2}
\label{GOCO}
%G_O &=&
%\frac{1}{\sqrt{10}}\left(1+\frac{2}{3N^2}\right)^{-1/2}\,,\,\,\,\, C_O
%=1
\ee We may then choose this operator as the first vector in the orthonormal
free basis (\ref{mix1}) and the remaining three basis vectors can be
constructed as linear combinations of the fields in (\ref{mix9}) that are
orthogonal to (\ref{O1def}). However, as already explained, when moving on to
the perturbative evaluation of two-point functions, we expect these operators
to mix non-trivially. Therefore, we are forced to evaluate all the ten
different two-point functions among them and solve the mixing problem.

To perform quantum calculations, we find it convenient to work in a ${\cal
N}=1$ setup where the field content of the theory is given in terms of one real
vector superfield $V$ and three chiral superfields $\Phi^i$ containing the six
scalars organized into the ${\bf 3}\times \bf{ \bar 3}$ of $SU(3) \subset
SU(4)$: \be \Phi^i\big|_{\theta=0}=\phi^i + i \phi^{i+3}~~~~~~~~
\Phib_i\big|_{\theta=0}=\phi^i - i \phi^{i+3} ~~~~~~~~~~i=1,\ldots,3 \ee The
classical action is (we use the notations of
\cite{Penati:1999ba,Penati:2000zv,Penati:2001sv}) \bea S &=&\int d^8z~ {\rm
Tr}\left(e^{-gV} \Phib_i e^{gV} \Phi^i\right)+ \frac{1}{2g^2}
\int d^6z~ {\rm Tr} W^\a W_\a\nonumber\\
&&+\frac{ig}{3!} {\rm Tr} \int d^6z~ \e_{ijk} \Phi^i
[\Phi^j,\Phi^k]+\frac{ig}{3!} {\rm Tr} \int d^6\bar{z}~ \e^{ijk}
\Phib_i [\Phib_j,\Phib_k] \label{actionYM}
\eea
where $W_\a=
i\Db^2(e^{-gV}D_\a e^{gV})$, and $V=V^aT^a$, $\Phi^i=\Phi^{ia}
T^a$, $T^a$ being $SU(N)$ matrices in the fundamental
representation. Since the theory is at its conformal fixed point,
possible divergences can arise only in correlators of composite
operators.

In terms of the (anti)chiral ${\cal N}=1$ superfields the four free operators
given in (\ref{mix9}) are\footnote{To be more precise, the operators
(\ref{mix9}) are the lowest components of these superfields.
In a small abuse of language we give  the
same names both to the superfields and to their lowest components.}
\bea
&& {\cal A}_1 = \frac{1}{2}\left[{\rm Tr}\left(\Phi^i\Phi^j\right)
{\rm Tr}\left(\Phib_i\Phib_j\right) + {\rm Tr}\left(\Phi^i\Phib_j\right)
{\rm Tr}\left(\Phi^j\Phib_i\right)\right]
\non\\
&& {\cal A}_2 = {\rm Tr}\left(\Phi^i\Phib_i\right)
{\rm Tr}\left(\Phi^j\Phib_j\right)
\non\\
&& {\cal A}_3 = {\rm Tr}\left(\Phi^i\Phi^j\Phib_i\Phib_j\right)
\non\\
&& {\cal A}_4 = \frac{1}{4}\left[
{\rm Tr}\left(\Phi^i\Phib_i\Phi^j\Phib_j\right) +
2 {\rm Tr}\left(\Phi^i\Phi^j\Phib_j\Phib_i\right) +
{\rm Tr}\left(\Phib_i\Phi^i\Phib_j\Phi^j\right)
\right]
\label{treeop}
\eea
When we turn on the interaction, in order to guarantee gauge invariance,
we have to insert extra exponential
factors inside the traces. The gauge invariant superfields are then given by
\bea
&& {\cal A}_1 = \frac{1}{2}\left[{\rm Tr}\left(\Phi^i\Phi^j\right)
{\rm Tr}\left(\Phib_i\Phib_j\right) +
{\rm Tr}\left(e^{gV}\Phi^i e^{-gV}\Phib_j\right)
{\rm Tr}\left(e^{gV}\Phi^j e^{-gV} \Phib_i\right)\right]
\non\\
&& {\cal A}_2 = {\rm Tr}\left(e^{gV}\Phi^i e^{-gV}\Phib_i\right)
{\rm Tr}\left(e^{gV}\Phi^j e^{-gV}\Phib_j\right)
\non\\
&& {\cal A}_3 = {\rm Tr}\left(e^{gV}\Phi^i\Phi^j e^{-gV}\Phib_i\Phib_j\right)
\non\\
&& {\cal A}_4 = \frac{1}{4}\left[
{\rm Tr}\left(e^{gV}\Phi^i e^{-gV}\Phib_ie^{gV}\Phi^j e^{-gV}\Phib_j\right) +
2 {\rm Tr}\left(e^{gV}\Phi^i\Phi^j e^{-gV}\Phib_j\Phib_i\right) \right.
\non\\
&& ~~~~~~~~~~~~~~~~~~~ \left. + {\rm Tr}\left(e^{-gV}\Phib_i e^{gV}
\Phi^i e^{-gV}\Phib_j e^{gV}\Phi^j\right)
\right]
\label{ginvop}
\eea

To compute correlation functions for these superfield operators, we introduce
the Euclidean generating functional \beq W[\vec{J}]= \int {\cal D}\Phi~{\cal
D}\Phib~{\cal D}V~e^{S + \int d^8z J^i {\cal A}_i} \label{genfunc} \eeq for
$n$--point functions \beq \langle {\cal A}_{i_1}(z_1) \cdots {\cal
A}_{i_n}(z_n) \rangle ~=~ \left. \frac{\d^n W}{\d J^{i_1}(z_1) \cdots \d
J^{i_n}(z_n)} \right|_{\vec{J}=0} \label{defcorr} \eeq where $z \equiv
(x,\theta, \bar{\theta})$. To evaluate perturbative contributions to $n$--point
functions it is sufficient to determine the contributions to $W[\vec{J}]$ at
order $n$ in the sources. In particular, for two-point functions we need
evaluate the quadratic terms \beq W[\vec{J}]\rightarrow \int d^4x_1~d^4x_2~
d^4\theta~ J^i(x_1,\theta,\bar{\theta}) \,
\frac{\rho_{ij}(g^2,N)}{[(x_1-x_2)^2]^{\Delta_0 + \g}} \, {\cal P}(D_\a,
\bar{D}_{\dot{\a}}) J^j(x_2,\theta,\bar{\theta}) \label{twopoint} \eeq where
${\cal P}(D_\a, \bar{D}_{\dot{\a}})$ is an operatorial expression built up from
spinorial derivatives. As discussed in the previous Section, the particular
dependence on the bosonic coordinates is fixed order by order by conformal
invariance and brings to the determination of the anomalous dimensions. The
only freedom left is in the appearance of possible perturbative multiplicative
corrections encoded in the matrix  $\rho_{ij}(g^2,N)$.

In order to compute the first-order contributions to (\ref{twopoint}), we first
draw all possible diagrams with two external sources at first order in $\l
\equiv {g^2N}/{4\pi^2}$. In the Feynman gauge, the relevant propagators read
(the ghosts do not contribute at this order) \be \langle V^a(z) V^b(0) \rangle
= - \frac{\d^{ab}}{4\pi^2} \frac{1}{x^2} \, \d^{(4)}(\theta) \qquad \langle
\Phi^{ia}(z) \bar{\Phi}^b_j(0) \rangle = \frac{\d^{ab}\, \d^i_j}{4\pi^2}
\frac{1}{x^2} \, \d^{(4)}(\theta) \ee while the vertices we need from the
action are

\bea
&& V_1 =ig f_{abc}\d^i_j \Phib^a_i V^b \Phi^{jc} \nonumber \\
&& V_2 = - \frac{g}{3!} \e_{ijk} f_{abc} \Phi^{ia} \Phi^{jb}
\Phi^{kc} \qquad \qquad \qquad \bar{V}_2 = - \frac{g}{3!} \e^{ijk}
f_{abc} \Phib_i^a \Phib_j^b \Phib_k^c \label{vertices}
\eea
with
additional $\Db^2$, $D^2$ factors for chiral, antichiral lines
respectively.

Our basic conventions to deal with color structures are as follows. For a
general simple Lie algebra we have \beq [T^a, T^b] ~=~ i f^{abc} T^c
\label{algebra} \eeq where $T^a$ are the generators  and $f^{abc}$ the
structure constants. The matrices $T^a$ are normalized as \beq {\rm Tr}(T^a
T^b) ~=~\d^{ab} \label{norm} \eeq We specialize to the case of $SU(N)$ Lie
algebra whose generators $T^a$, $a= 1, \cdots, N^2-1$ are taken in the
fundamental representation, i.e. they are $N \times N$ traceless matrices. The
basic relation which allows to deal with products of $T^a$'s is the following
\beq T^a_{pq} T^a_{rs}= \left( \d_{ps}\d_{qr}-
\frac{1}{N}\d_{pq}\d_{rs}\right). \label{Tcontract} \eeq {}From this identity,
together with (\ref{algebra}), we can easily obtain all the identities used to
compute the color structures associated to the Feynman diagrams relevant for
the two--point correlation functions.

As a second step one needs to perform the superspace $D$--algebra to reduce
each diagram to an ordinary multiloop integral. To deal with possible
divergences in these integrals we perform massless dimensional regularization
by analytically continuing the theory to $d$ dimensions, $d=4-2\e$. We evaluate
the integrals using the general formula \be \int d^d x
\frac{1}{(x^2)^a[(x-y)^2]^b} = \pi^2 \frac{\G(a+b-\frac{d}{2} )}{\G(a)\G(b)}
\frac{\G(\frac{d}{2}-a) \G(\frac{d}{2} -b)} {\G(d-a-b)}
\frac{1}{(x^2)^{a+b-\frac{d}{2}}} \label{int1} \ee and extracting the
$\frac{1}{\e}$ pole. The leading behavior of the integrals we use in the
calculation is \be \int d^d x \frac{1}{(x^2)^{1-\e}[(x-y)^2]^{2-2\e}} \sim
\frac{\pi^2}{\e} \frac{1}{x^2} \label{int2} \ee \be \int d^d x
\frac{1}{(x^2)^{2-2\e}[(x-y)^2]^{2-2\e}} \sim 2 \, \frac{\pi^2}{\e}
\frac{1}{(x^2)^2} \label{int3} \ee

\section{One-loop anomalous dimension matrix
  and quasiprimary composite operators in ${\cal N}=4$ SYM}

In this section we compute the full one-loop anomalous dimension matrix for the
operators (\ref{mix9}) using the ${\cal N}=1$ superspace approach described in
the preceding section. Let us first write the four gauge invariant superfields
(\ref{ginvop}) by expanding the exponentials inside the traces up to order $gV$
\begin{eqnarray}
\label{expop} && {\cal A}_1 = \frac{1}{2}\Phi^{ia} \Phi^{jb}
\Phib_i^c \Phib_j^d \left[(\d_{ab}\d_{cd}+\d_{ad}\d_{bc}) + {\rm
i}gV^e (f^{dea}\d^{bc}
+f^{ceb}\d^{ad}) \right] \nonumber\\
&& {\cal A}_2 = \Phi^{ia} \Phi^{jb} \Phib_i^c \Phib_j^d
\left[(\d_{ac}\d_{bd} + {\rm i}gV^e (f^{cea}\d^{bd}
+f^{deb}\d^{ac}) \right] \nonumber\\
&& {\cal A}_3 = \Phi^{ia} \Phi^{jb} \Phib_i^c \Phib_j^d \left[{\rm
Tr}(T_a T_b T_c T_d) + {\rm i}gV^e \left(f^{gea} {\rm Tr}(T_g T_b
T_c T_d) + f^{heb}{\rm Tr}(T_a T_h T_c T_d)\right) \right]
 \nonumber\\
&& {\cal A}_4 = \frac{1}{4}\Phi^{ia} \Phi^{jb} \Phib_i^c \Phib_j^d
\Big\{ {\rm Tr}(T_a T_c T_b T_d) + 2 {\rm Tr}(T_a T_b T_d T_c)
+ {\rm Tr}(T_a T_d T_b T_c)\nonumber\\
&&~~~~~~~ + {\rm i}gV^e \left[f^{gea}\left({\rm Tr}(T_g T_c T_b T_d) +
2 {\rm Tr}(T_g T_b T_d T_c) + {\rm Tr}(T_g T_d T_b T_c)\right)\right.
\nonumber\\
&&~~~~~~~~~~~~~~~ + \left.f^{heb}\left({\rm Tr}(T_a T_c T_h T_d) +
2 {\rm Tr}(T_a T_h T_d T_c)
+ {\rm Tr}(T_a T_d T_h T_c)\right)\right]\Big\}
\end{eqnarray}
First of all, we are interested in the calculation of the tree-level two-point
functions of these operators. Neglecting for the moment their color and flavor
structures, we concentrate on the two-point function of the generic operator
$\Phi\Phi\Phib\Phib$. Its diagram is shown in Figure 1 where the superspace
derivatives acting on the propagators are explicitly indicated. In this case
the D-algebra is performed rather easily and the result is given by
\begin{equation}
W[J]_{tree}\rightarrow \int d^4x_1~d^4x_2~ d^4\theta~
J(x_1,\theta,\bar{\theta}) \, \frac{1}{(x_1-x_2)^8} \, \left(D^2 \Db^2 +
\frac{{\rm i}}{2}\pa_{\a\ad}D^{\a}\Db^{\ad} + \frac{1}{6}\square \right)
J(x_2,\theta,\bar{\theta}) \label{tree}
\end{equation}

\vskip 18pt
\noindent
%---------- FIGURE TOP ------------
\begin{minipage}{\textwidth}
\begin{center}
\includegraphics[width=0.60\textwidth]{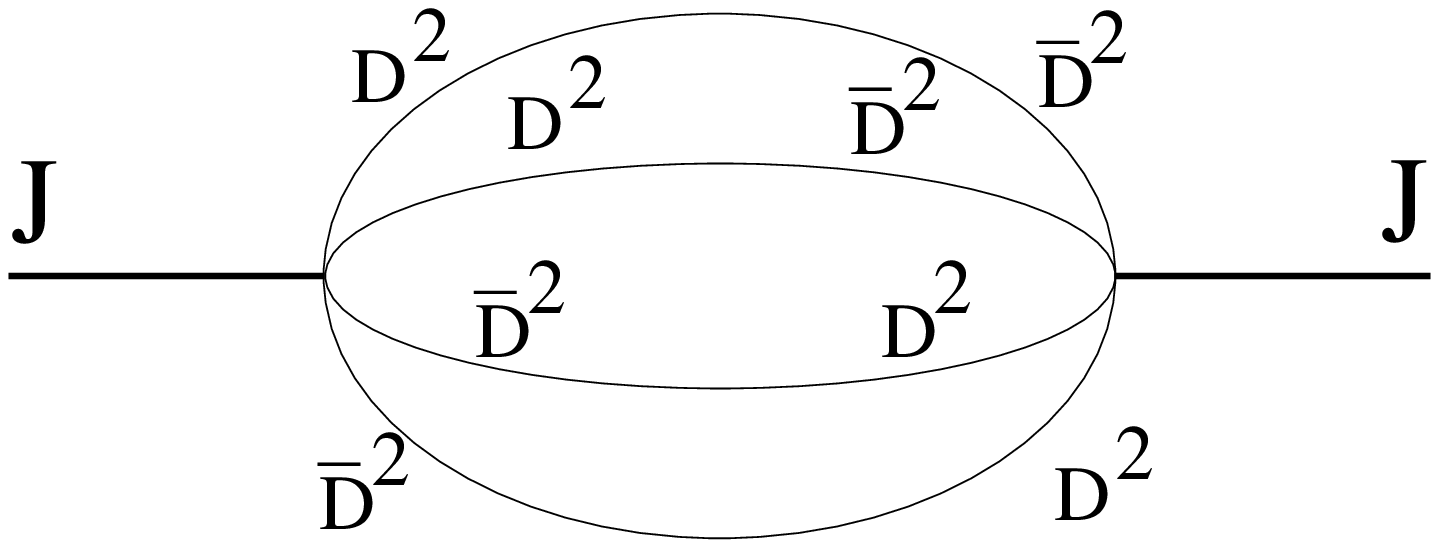}
\end{center}
\begin{center}
{\small{Figure 1: Tree level contribution to $\langle A(z_1)A(z_2)\rangle$}}
\end{center}
\end{minipage}
%---------- FIGURE END ------------

\vskip 20pt

To evaluate the full result for $\langle {\cal A}_i {\cal A}_j\rangle_0$ we
have to take into account the combinatorics coming from the sum over flavor
indices and contractions of the color tensor structures appearing in
(\ref{expop}). This is a straightforward algebraic calculation which finally
gives \bea \langle {\cal A}_i(z)  {\cal A}_j(0) \rangle_0 &=&
\frac{3(N^2-1)}{(4\pi^2)^4} \left( \matrix{ \frac{7 N^2 + 2}{2} & N^2 + 6 &
\frac{7N^2 - 8}{N} & \frac{9N^2 - 16}{2N}\cr N^2 + 6 & 2(3 N^2 - 2) &
\frac{2(N^2 - 4)}{N} & \frac{7N^2 - 8}{N}\cr \frac{7N^2 - 8}{N} & \frac{2(N^2 -
4)}{N} & \frac{3N^4 - 8N^2 + 24}{N^2} & \frac{N^4 - 16N^2 + 48}{2N^2}\cr
\frac{9N^2 - 16}{2N} & \frac{7N^2 - 8}{N} & \frac{N^4 - 16N^2 + 48}{2N^2} &
\frac{7N^4 - 32N^2 + 96}{4N^2}
}\right) \nonumber \\
&&~~~~~\nonumber \\
&&\hspace{2cm}\times\left(D^2 \Db^2 + \frac{{\rm
i}}{2}\pa_{\a\ad}D^{\a}\Db^{\ad} + \frac{1}{6}\square \right)\frac{1}{x^8}
\d^{(4)}(\theta) \label{treematrix} \eea We notice that the operators ${\cal
A}_i$ do not form an orthogonal basis at tree level. Diagonalizing the previous
matrix one can find a suitable orthonormal basis. This is discussed in details
in Appendix B where the diagonalization is performed for large $N$.

Let us now move to the computation of the one-loop two-point function. The
various contributions are shown in Figure 2. To draw the diagrams 2a, 2b, 2c
and 2f, 2g, 2h we have used only the interaction vertices given in
(\ref{vertices}), while diagrams 2d and 2e contain also external vertices with
one vector field, as given in (\ref{expop}).

\vskip 18pt \noindent
%---------- FIGURE TOP ------------
\begin{minipage}{\textwidth}
\begin{center}
\includegraphics[width=0.80\textwidth]{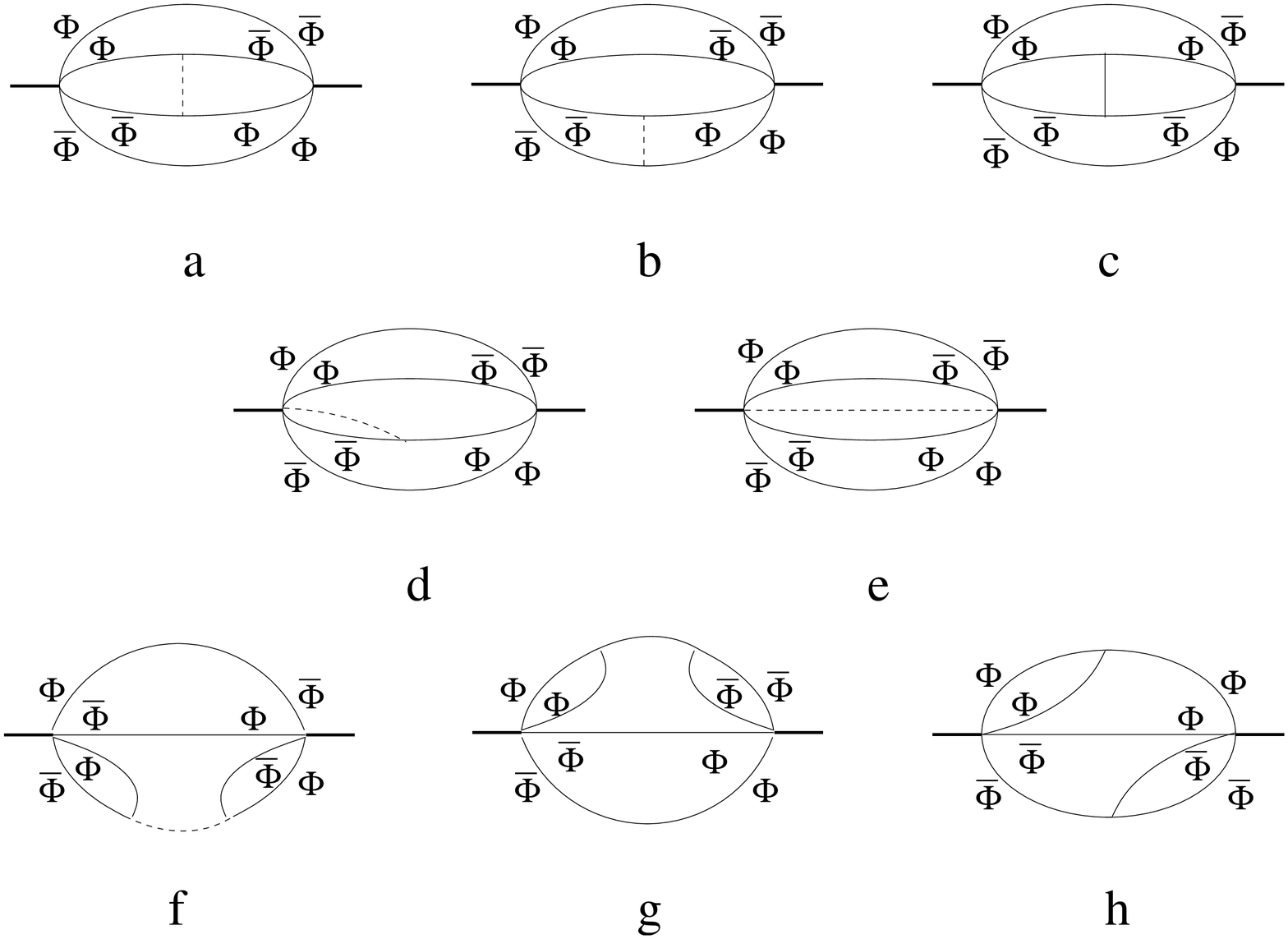}
\end{center}
\begin{center}
{\small{Figure 2: One--loop contributions to $\langle A(z_1)A(z_2)\rangle$}}
\end{center}
\end{minipage}
%---------- FIGURE END ------------

\vskip 20pt

Since we are interested merely in the anomalous dimensions of the operators, in
accordance with (\ref{mix2}) we consider only the logarithmic divergences
which, in dimensional regularization, appear as ${1}/{\epsilon}$ poles
\footnote{Note that, in a momentum space approach, as the one used in
\cite{Penati:2001sv}, these terms are instead the ones appearing as
${1}/{\epsilon^2}$ poles.}. Therefore, we only concentrate on ${1}/{\epsilon}$
divergent diagrams. After performing the $D$--algebra and looking at the
structure of the integrals which arise, it is not difficult to see that graphs
2a, 2c, 2d, 2f, and 2g are the only divergent diagrams, whereas graphs 2b, 2e
and 2h give only finite contributions which contribute to the $\rho$ matrix as
defined in (\ref{mix2}).

After $D$--algebra, the structure of superspace derivatives we are left with is
the same as the one at tree-level (see eq. (\ref{tree})). Along the
calculation, we can then concentrate only on the contributions proportional to
$D^2\Db^2$, the other ones following by supersymmetry. The graphs 2a, 2c and 2d
then reduce to the ordinary Feynman diagram in Figure 3a, while the reduction
of graphs 2f and 2g produce the diagram in Figure 3b. The corresponding
integrals are (\ref{int2}) and (\ref{int3}), respectively. They both contribute
with a simple pole divergence, the integral of Figure  3b giving twice the
result of the one in Figure  3a.

\noindent
%---------- FIGURE TOP ------------
\begin{minipage}{\textwidth}
\begin{center}
\includegraphics[width=0.70\textwidth]{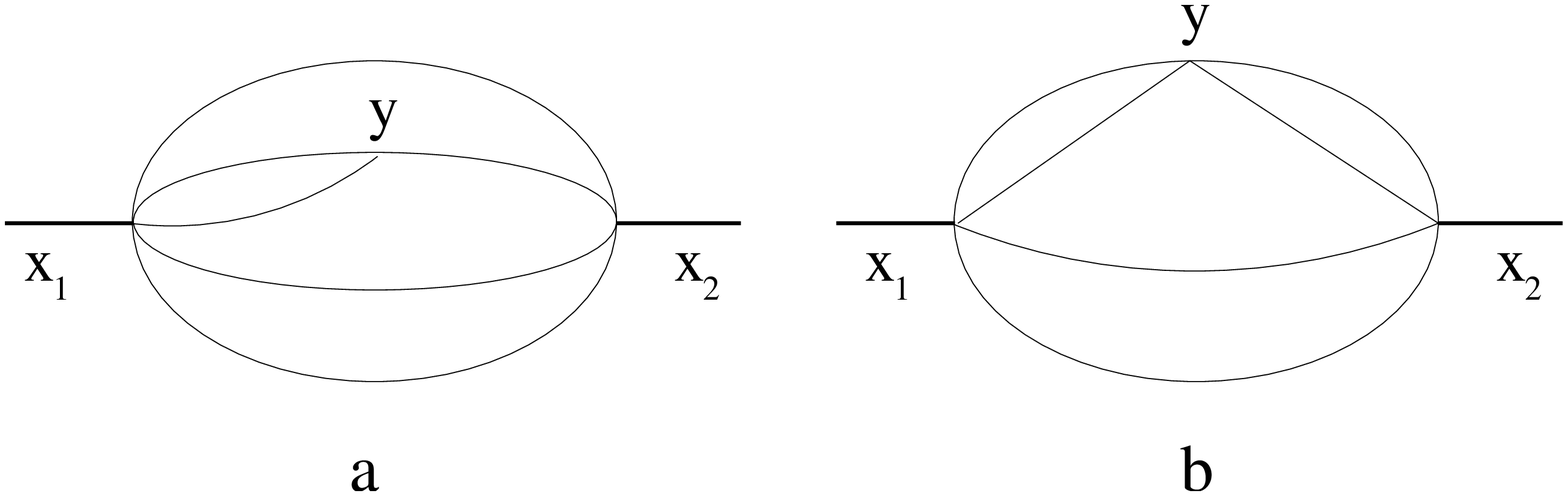}
\end{center}
\begin{center}
{\small{Figure 3: Bosonic integrals resulting after D--algebra}}
\end{center}
\end{minipage}
%---------- FIGURE END ------------

\vskip 20pt

At this point we have to take into account the factors coming from the flavor
and color combinatorics. This is the most tedious and tiring part of the
calculation, even if rather straightforward. To perform the calculation we have
first done the flavor contractions by hand, while to deal with the color
contractions we have taken advantage of a computer program we have implemented
with {\em Mathematica} \cite{math}. As a result, the anomalous dimension matrix
of the four operators ${\cal A}_1, \ldots, {\cal A}_4$ is given by \bea
\label{loopmatrix}
&&~~~~~~~\nonumber \\
&& -\frac{3}{2}\lambda(N^2-1)
\left(
\matrix{
-2N^2 + 13 & -6(2N^2 + 7) & \frac{21N^2 + 16}{N} & -\frac{53N^2 - 32}{2N}\cr
-6(2N^2 + 7) & -12(6N^2 + 1) & \frac{6(N^2 + 16)}{N} & -\frac{3(33N^2 - 32)}
{N}\cr
\frac{21N^2 + 16}{N} & \frac{6(N^2 + 16)}{N} & -\frac{11N^4 - 96N^2 + 128}
{N^2} & \frac{3N^4 + 112N^2 - 256}{2N^2}\cr
-\frac{53N^2 - 32}{2N} & -\frac{3(33N^2 - 32)}{N} & \frac{3N^4 + 112N^2 - 256}
{2N^2} & -\frac{59N^4 - 64N^2 + 512}{4N^2}
}
\right) \nonumber \\
&&~~~~~~~ \eea {}From the results (\ref{treematrix}, \ref{loopmatrix}), we see
that the four operators do mix non-trivially at tree and one--loop level. As
discussed in section 3.2, we need  to solve the mixing in order to find the
anomalous dimensions of the quasiprimary operators belonging to the singlet
sector of the theory. This amounts to building first of all an orthonormal
basis at tree level out of the four operators ${\cal A}_1, \ldots, {\cal A}_4$,
and then to writing the anomalous dimension matrix in this new basis. It is
from this matrix (once we have diagonalized it) that we can read off the pure
conformal operators of the theory and the values of their anomalous dimensions
up to order $g^2$.

The procedure is described in some details in Appendix B, where we have
computed the values of the anomalous dimensions up to order ${1}/{N^2}$, and
have identified the corresponding quasiprimary operators $\S_1, \ldots, \S_4$.
The explicit expressions for the latter are given in (\ref{eigenvectors1}) and
their corresponding anomalous dimensions are
\bea \label{anom}
\gamma_1 &=& -10\frac{\lambda}{N^2}\nonumber\\
\gamma_2 &=& \left(6 + \frac{20}{N^2}\right) \lambda \nonumber\\
\gamma_3 &=& \left[\frac{13 + \sqrt{41}}{4} -
\frac{5}{41N^2}(41+19\sqrt{41})\right] \lambda \nonumber\\
\gamma_4 &=&\left[\frac{13 - \sqrt{41}}{4} -
\frac{5}{41N^2}(41-19\sqrt{41})\right] \lambda
\eea
In terms of these
operators, the (normalized) operator ${\cal O}_1$ appearing in the OPE
(\ref{OPEO1}) is
\be \label{O1pure} {\cal O}_1 = \S_1 -
\frac{1}{N}\left[\a_3\S_3 + \a_4 \S_4\right]
+\frac{1}{N^2}\left[-\frac{21}{16}\S_1 + \frac{3\sqrt{5}}{4}\S_2 \right] +
O\left(\frac{1}{N^3}\right)
\ee
where $\a_3$ and $\a_4$ are given in
(\ref{agamma}). Therefore, at leading order in $N$, the operator which appears
in the singlet channel of the OPE of the two CPOs ${\cal O}_{\bf 20}$ is a
quasiprimary operator with vanishing anomalous dimension. It becomes a mixture
of pure states when subleading corrections in $N$ are taken into account. The
meaning of this result will be discussed on general grounds in the following
Section.

To conclude this Section, we check our result against the four-point OPE
analysis mentioned in Section 3.3 and discussed in detail in Appendix A. From
(\ref{O1pure}) we see that the condition (\ref{cons1}) is satisfied up to order
${1}/{N^2}$:

\be 1-\frac{21}{8N^2}+\frac{1}{N^2}\left[\a_3^2 + \a_4^2 \right]=1 +
O\left(\frac{1}{N^3} \right) \ee while by virtue of the results (\ref{anom}),
the condition (\ref{cons2}) reads \be \label{cons22} -\frac{10\lambda}{N^2}
+\frac{1}{N^2}\left[\a_3^2 \gamma_3 +\a_4^2
\gamma_4\right]=-\frac{4\lambda}{N^2} + O\left(\frac{1}{N^3} \right) \ee which
can also holds at order $\frac{1}{N^2}$.
\\
Finally, condition (\ref{cons3}) is also satisfied at this same order as can be
easily checked \be \label{cons33} \frac{1}{N^2}\left[\a_3^2 \gamma_3^2 + \a_4^2
\gamma_4^2\right]= 18\frac{\lambda^2}{N^2} + O\left(\frac{1}{N^3} \right). \ee

\section{Non-protected operators and multi--particle states in supergravity}

In the previous two sections we have studied in detail the four possible scalar
quasiprimary operators which are $SU(4)$ singlets of canonical dimension four
in ${\cal N}=4$ SYM. At one loop, we identified them explicitly in terms of the
elementary superfields (\ref{eigenvectors1}) whose anomalous dimensions are
given in (\ref{anom}).

The first important observation that follows from our results is connected with
the anomalous dimensions (\ref{anom}). We see that one of the four anomalous
dimensions, $\gamma_1$, is suppressed by a factor of $1/N^2$ with respect to
the remaining three, $\gamma_{2,3,4}$. Although this is merely a one-loop
result, it points at a fundamental difference between the quasiprimary operator
$\S_1$ on the one hand, and $\S_{2,3,4}$ on the other. Namely, the anomalous
dimensions of the latter three operators behave like the anomalous dimensions
of the $K$-class operators, that is, they are positive and of order
$O(\lambda)$. Therefore, it is justified to believe that these three operators
are dual to string modes. Their anomalous dimensions become infinite when
$\lambda, N \rightarrow \infty$, so that they decouple in the supergravity
limit. On the other hand, the anomalous dimension of the operator $\S_1$, being
of order $1/N^2$ at one loop, has a chance to remain finite in the supergravity
limit.

The above observations are in agreement with the known strong coupling
calculations of the four-point function of the CPOs (\ref{O20}). In particular,
in \cite{Arutyunov:2000ku} it was shown that the contribution of {\it all} the
operators in the $K$-class is absent in the supergravity limit of that
four-point function. However, the contribution of scalar fields with canonical
dimension four was shown to be non-trivial. Our result above indicates that
{\it only one} of the four possible quasiprimary operators with canonical
dimension four might survive in the supergravity limit and give a non-trivial
contribution to the four-point function of the CPOs (\ref{O20}). This is the
quasiprimary operator $\S_1$ which therefore should correspond to a
two-particle supergravity bound state.

Further support for our interpretation of $\S_1$ as a two-particle supergravity
bound state comes from the usual identification of the conformal anomalous
dimension with the energy of a state in a radial quantization (see e.g.
\cite{Witten:1998qj}). A  negative anomalous dimension would then correspond to
a {\it binding energy}. In this sense, it is quite suggestive for the nature of
the state corresponding to the quasiprimary operator $\S_1$ that both its
one-loop as well as its strong coupling anomalous dimensions are negative. It
would be interesting to find out if this property persists at higher orders in
perturbation theory.

Our next observation concerns the operator $\Sigma_2$. {}From (\ref{anom}) we
see that in the large $N$ limit its anomalous dimension $\gamma_2=6\lambda$ of
this operator equals twice the one-loop anomalous dimension
$\gamma_{\cal K}=3\lambda$
of the Konishi scalar ${\cal}\sim{\rm Tr}\left(e^{gV}\Phi^i e^{-gV}\Phib_i\right)$
\cite{Anselmi,Bianchi:1999ge,Arutyunov:2000im}. Alternatively, one can also
deduce from (\ref{Konishisquared}) that for large $N$ $\Sigma_2={\cal K}^2
\sim  {\cal A}_2$, i.e.,  the operator $\Sigma_2$ coincides
with the canonically normalized
square of the Konishi scalar. In other words, we observe that in the large $N$
limit the anomalous dimensions of the individual constituents of the composite
operator $\Sigma_1$ just add up. Owing to this property it is then natural to
identify $\Sigma_2$ with an operator dual to a two-particle string state.

Another intriguing consequence of our results comes when we compare our formula
(\ref{O1pure}) and the free OPE (\ref{OPEO1}). In principle, there is no reason
why the free operator ${\cal O}_1$ that appears in the OPE (\ref{OPEO1}) and
the quasiprimary operator $\S_1$  might bare any resemblance. Nevertheless, it
is easily seen from (\ref{O1pure}) or from (\ref{eigenvectors1}) and
(\ref{GOCO}) that in the large $N$ limit ${\cal O}_1$ coincides with the
quasiprimary operator $\S_1$. Indeed, to calculate the two-point function of
${\cal O}_1$ one should use the decomposition (\ref{O1pure}) into quasiprimary
operators. However, as we have argued above, we expect that in the supergravity
limit the operators $\S_{2,3,4}$ decouple since they are dual to massive string
modes. Then, since we are at the same time considering the large $N$ limit, we
see from (\ref{O1pure}) and (\ref{eigenvectors1}) that ${\cal O}_1$ coincides
with the quasiprimary operator $\S_1$ dual to a two-particle supergravity state
in the supergravity limit. It is remarkable that the OPE itself somehow {\it
knows} the special r\^ole of the operator ${\cal O}_1$ and gives us its
explicit expression in terms of elementary superfields without the need to go
through the process of diagonalization of all the possible mixed fields.

Now we want to argue that that the suppression of the anomalous dimension of
$\S_1$ should work to all orders in perturbation theory. This is due to the
special nature of the operator ${\cal O}_1$ which can be written as a direct
product of two CPOs: \be \label{O1CPO} {\cal O}_1(x) = {\cal O}_{\bf
20}^{LM}(x){\cal O}_{\bf 20}^{LM}(x)\,. \ee One can show that all of its
two-point functions, with itself and with the remaining three operators ${\cal
A}_{2,3,4}$ from (\ref{mix9}) give logarithmic terms that are at least $1/N$
suppressed.

Let us first discuss the two-point function $\langle {\cal O}_1 {\cal O}_1
\rangle$. From Figure 4 we see that at tree level it includes a connected and a
disconnected parts. The connected part is $1/N^2$ suppressed as it is {\it
non-planar}.

\setcounter{figure}3

\begin{figure}[ht]
\centering \epsfysize=8cm
\begin{minipage}{14cm}

\includegraphics[width=10cm]{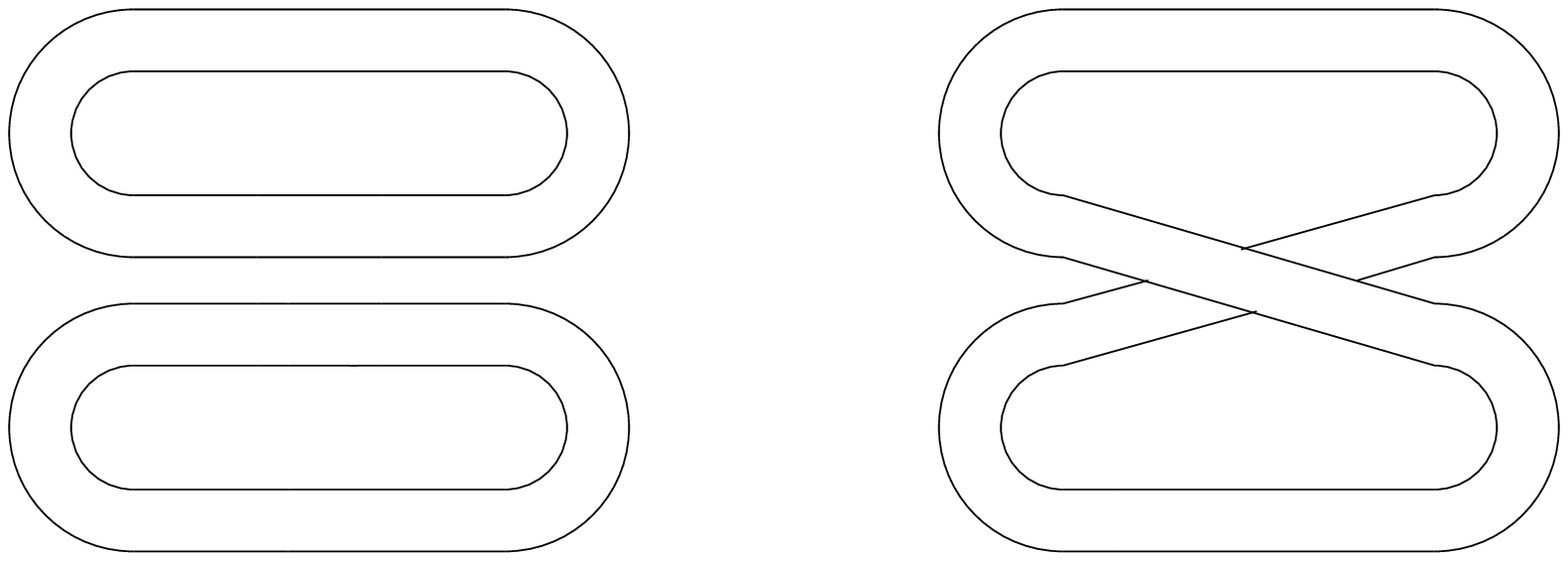}
\centering \caption{ Contributions to the tree-level two-point function of
  ${\cal O}_1$ in
    the double line notation.}
\end{minipage}
\end{figure}

The perturbative corrections to the above picture can be of three types (Figure
5). Firstly, there can be corrections to {\it each} of the disconnected graphs
separately. Secondly, there can be lines that {\it connect the disconnected}
graphs. Finally, there can be perturbative corrections to the {\it non-planar}
connected graphs. Among them, the leading contribution to the $1/N$ expansion
is supplied by the first type of graphs, where all the corrections are confined
to the disconnected parts. Now, the crucial point is that these graphs can be
factorized as follows:
\begin{equation}\label{factgr}
  \langle {\cal O}_{1}(x) {\cal O}_{1}(y) \rangle^{\rm disc.}_1 =
  2\,\langle {\cal O}^{LM}_{\bf 20}(x) {\cal O}^{NP}_{\bf 20}(y)\rangle_0\
  \langle {\cal O}^{LM}_{\bf 20}(x) {\cal O}^{NP}_{\bf 20}(y)\rangle_1\,,
\end{equation}
where the subscript $\langle\rangle_{0,1}$ indicates the perturbation level.
For this subset the trace over the $SO(6)$ indices in (\ref{O1CPO}) is of no
relevance, so we can treat each factor in the right-hand side of (\ref{factgr})
as an independent two-point function of the  {\it protected operator} ${\cal
O}_{\bf 20}$. The protected operators have a fixed scaling dimension determined
by their nature of short superconformal representations. Consequently, all such
divergent\footnote{In fact, the 1/2 BPS short operator ${\cal O}_{\bf 20}$ is
completely non-renormalized, including the finite quantum corrections. However,
here we are only interested in logarithmic terms which come from the divergent
graphs.} graphs, which could possibly contribute to the anomalous dimension of
${\cal O}_{\bf 20}$, should add up to zero. All the remaining {\it planar} and
{\it non-planar connected} graphs are subleading, at most of order $1/N^2$.

Clearly, this mechanism works to {\it all orders} in perturbation theory.
Indeed, at order $n$ eq. (\ref{factgr}) can be replaced by
\begin{equation}\label{factgrn}
  \langle {\cal O}_{1}(x) {\cal O}_{1}(y) \rangle^{\rm disc.}_n =
  \sum^n_{m=0} C_{nm}\,\langle {\cal O}^{LM}_{\bf 20}(x) {\cal O}^{NP}_{\bf 20}(y)\rangle_{m}\
  \langle {\cal O}^{LM}_{\bf 20}(x) {\cal O}^{NP}_{\bf 20}(y)\rangle_{n-m}\,.
\end{equation}
In this sum, for a given perturbative correction of order $m$ to the first
disconnected part we have collected all the corrections of order $n-m$ to the
second disconnected part. The latter correspond to the protected two-point
function at order $n-m$, so they give no logarithmic terms. The same argument
can then be applied to the first part of order $m$.

\begin{figure}[ht]
\centering \epsfysize=8cm
\begin{minipage}{14cm}

\includegraphics[width=10cm]{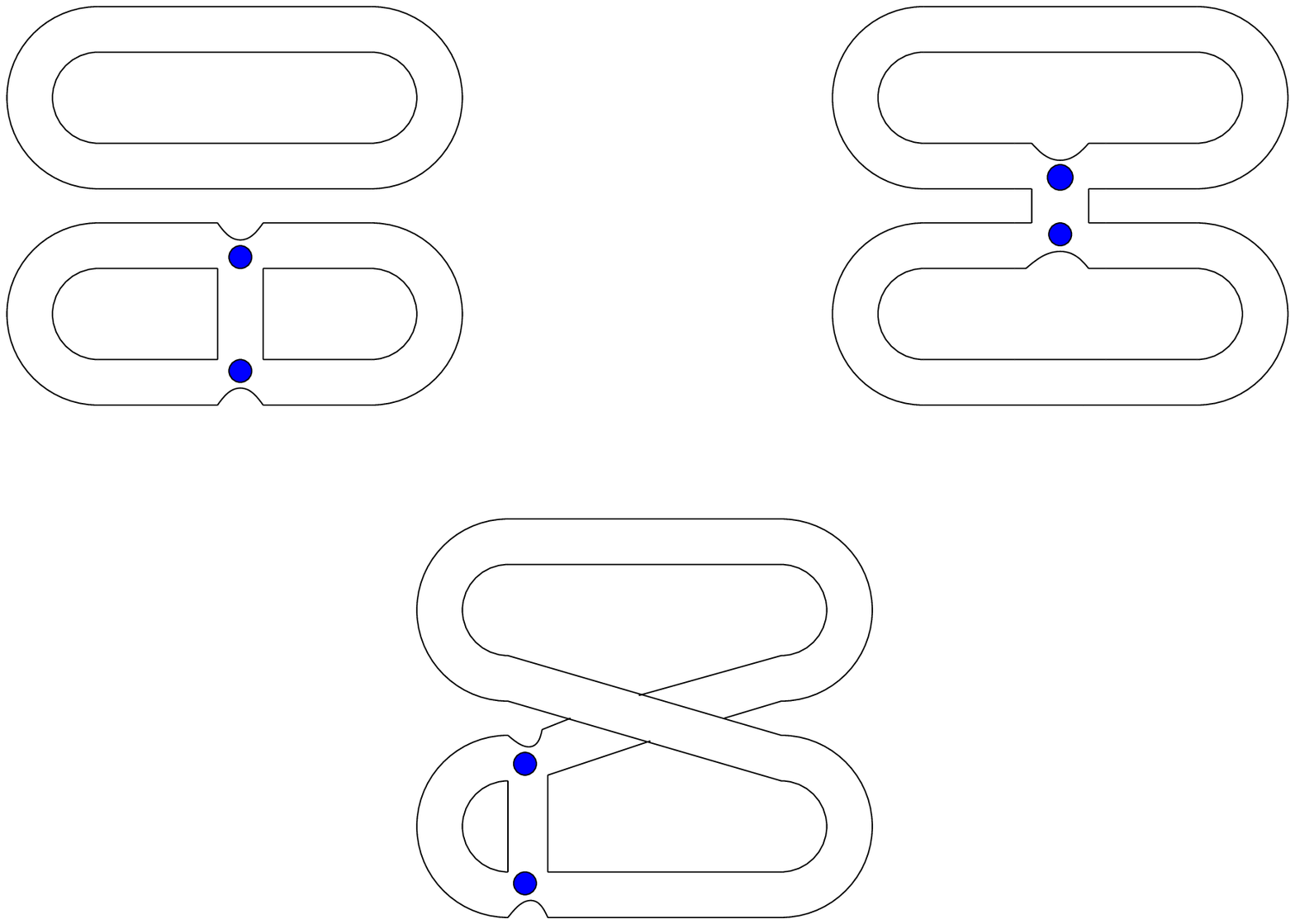}
\centering \caption{ Contributions to the one-loop two-point function of
  ${\cal O}_1$ in
    the double line notation.}

\end{minipage}
\end{figure}

It remains to show that the two-point functions of ${\cal O}_1$ with the
operators ${\cal A}_{2,3,4}$ from (\ref{mix9}) give logarithmic terms that are
at least $1/N$ suppressed. In the case of ${\cal A}_{3,4}$ this is due to the
fact that they are single-trace operators, so we have one color trace fewer in
the graphs. The disconnected (leading) sector of the two-point function of the
double-trace operators ${\cal O}_1 \sim {\cal O}_{\bf 20}{\cal O}_{\bf 20}$ and
${\cal A}_{2} \sim {\cal K}{\cal K}$ is empty, since it is not possible
to factorize it into
two-point functions matching an ${\cal O}_{\bf 20}$ factor with a singlet
${\cal K}$
factor (see Figures 4 and 5). Again, this is true to all orders in perturbation
theory. Thus, all the entries in the first row of the anomalous dimension
matrix (\ref{admatrix}) are at least $1/N$ suppressed. When diagonalizing such
a matrix in the large $N$ limit, the operator ${\cal O}_1$ effectively becomes
the eigenvector $\S_1$ with a subleading eigenvalue. This is the mechanism
which is responsible for the $1/N^2$ suppression of the anomalous dimension of
$\S_1$ and thus for the existence of two-particle supergravity states with
non-zero binding energy.

It is important to mention that the same mechanism can be applied to any
multitrace operator which is defined as a product of any number of protected
operators, not necessarily identical. Then we may conjecture that such
operators constitute a substantial part, if not all, of the supergravity
multiparticle (scalar) states. Whether one can construct multiparticle states
with non-vanishing spin in this way is not immediately clear. The point is that
BPS short operators like ${\cal O}_{\bf 20}$ are always scalars. However, we
know another type of protected operators which can have arbitrary spin
\cite{Arutyunov:2001qw,Heslop:2001dr}. It is possible that they, together with
the BPS operators, play an important r\^ole in constructing multiparticle
states.

%%%%%%%%%%%%%%%%%%%%%%%%%%%%%%%%%%%%%%%%%%%%

%---------- FIGURE TOP ------------
%\begin{minipage}{\textwidth}
%\begin{center}
%\includegraphics[width=0.70\textwidth]{dline_tree1.eps}
%\end{center}
%\begin{center}
%{\small{Figure 5: Contributions to the one-loop two-point function of
%    ${\cal O}_1$ in
%    the double line notation. }}
%\end{center}
%\end{minipage}
%%---------- FIGURE END ------------
%\vspace{.8cm}

In conclusion, we have studied the dimension 4 scalar quadrilinear operators
that appear in the $SU(4)$ singlet channel of the OPE between two lowest weight
CPOs in ${\cal N}=4$ SYM. At one loop, due to the four-fold degeneracy at the
free level, we had to diagonalize a 4$\times$4 matrix of two-point functions.
We were able to identify the one-loop quasiprimary operators, to find their
anomalous dimensions and to give explicit expressions for them in terms of the
elementary superfields. We found that one of the four quasiprimary operators is
a possible candidate to correspond to a two-particle supergravity state, since
its anomalous dimension is $1/N^2$ suppressed with regards to the anomalous
dimensions of operators that are dual to string modes. We pointed out that the
special structure of the operator ${\cal O}_1$ is connected to the mechanism
responsible for such a $1/N^2$ suppression. Curiously, the free conformal OPE
seems to provide the explicit expression for the quasiprimary operator dual to
the two-particle supergravity state, to leading order in $N$.

Although our results are one-loop and restricted to the simplest scalar
quadrilinear operators, we believe that the general pattern we sketched remains
valid both at higher loops as well as for non-scalar quadrilinear operators. It
would be desirable to extend our results to two loops in order to get a clearer
picture. The mechanism for the $1/N^2$ suppression that we have revealed
affects {\it both planar and non-planar} graphs. It seems therefore possible
that non-planar graphs survive in the supergravity limit and play an important
r\^ole in the formation of multiparticle supergravity states. It would be
interesting to understand better the strong coupling supergravity limit that
seems to be more complicated than initially thought, in order to resolve the
puzzle of the multiparticle supergravity states.

We have discussed the general approach to solve the mixing and splitting of
operators. Our strategy works in any CFT and can be applied to the study of
composite operators in a variety of cases, such as ${\cal N}=2$  and ${\cal
N}=1$ superconformal theories, as well as in studies of multitrace operators in
the pp-wave limit. \bigskip

{\bf Note added} When our paper was ready for submission to the e-archive, we
noticed a new paper \cite{BERS} were the operator mixing problem is discussed
for scalar operators of approximate dimension 4 but belonging to the ${\bf 20}$
of $SU(4)$. Our operators are $SU(4)$ singlets.

{\bf Note added in proof} In addition to the four linearly independent
operators that we considered in the text, Eq. (4.1), there exist two more
operators with the same quantum numbers (apart from fermionic operators which
do not contribute to the two-point functions at level $O(\lambda)$). Their
lowest components are ${\rm Tr}[ D_\mu\phi^I D^\m\phi^I]$  and ${\rm
Tr}[F_{\m\n}F^{\m\n}]$. The first of them is the leading term in the conformal
descendant of the Konishi scalar $\square K$, and a combination of the two is
the leading term in the ${\cal N} = 4$ SYM Lagrangian which belongs to a
protected multiplet. Hence, these operators can be identified as pure states.
In addition, at tree level they are orthogonal to all four operators in Eq.
(4.1). Then it is easy to see that their presence can only modify the one-loop
pure state operators in Eq. (B.9) by terms of order $O(\lambda)$ which are
beyond the scope of this work.

\newpage

\section*{Acknowledgements}
\noindent

We thank A. Vicini for the help in using {\em Mathematica}. 
We acknowledge
very useful discussions and correspondence with B. Eden, 
E. Gardi, H. Osborn, R.
Roberts, W. R\"uhl, C. Sachrajda and I. Todorov.  E.S. is grateful to the
Theoretical Physics group for the warm hospitality at the University of
Milano-Bicocca, where part of this work was done.

This work has been supported in
part by INFN, MURST and the European Commission RTN program
HPRN--CT--2000--00131, in which
G. A. is associated to
the University of Bonn, S. P. is associated to
the University of Padova and A. S. is associated to the University of
Torino. The work of G. A. has been also supported by the DFG and
in part by RFBI grant
N99-01-00166.

%\newpage

\appendix

\section{Four-point functions and the OPE}

Here we recall the relevant facts about conformal partial wave expansions
required to carry out an independent verification of our results at the OPE
level (see \cite{Arutyunov:2000ku,Arutyunov:2000im,Arutyunov:2001mh} for
details) and show how to extract some information about the quasiprimary
operators $\Sigma_i$ encoded in the two-loop four-point function of $O_{\bf
20}^{LM}$.

Every irreducible representation of the R symmetry algebra  in the tensor
product decomposition (\ref{decomp}) appears in the OPE of $O_{\bf 20}^{LM}$ as
an infinite tower of operators $O_{\Delta,l}$, where $\Delta$ is the conformal
dimension and $l$ is the Lorentz spin. We are interested only in the singlet R
symmetry channel for which the corresponding contribution to the four-point
function can be viewed as an expansion of the type \bea \label{GE} F=\langle
O_{\bf 20}^{LM}(x_1)O_{\bf 20}^{LM}(x_2)O_{\bf 20}^{NP}(x_3)O_{\bf
20}^{NP}(x_4)\rangle = \sum_{\Delta, l}A_{\Delta, l}{\cal H}_{\Delta,
l}(x_1,x_2,x_3,x_4) \, . \eea Here ${\cal H}_{\Delta, l}(x_1,x_2,x_3,x_4)$ is
the canonically normalized conformal partial wave amplitude (CPWA) representing
the individual contribution (exchange) of an operator $O_{\Delta,l}$ and
$A_{\Delta, l}$ is a normalization coefficient. We treat the CPWA as a double
series of the type \bea \label{cpwa} {\cal H}_{\Delta,
l}(x_1,x_2,x_3,x_4)=\frac{1}{x_{12}^4x_{34}^4}v^{(\D-l)/2}
\sum_{n,m}c_{mn}^{\Delta,l}v^nY^m \, , \eea where
$Y=1-(x_{13}^2x_{24}^2)/(x_{14}^2x_{23}^2)$ is the other conformally invariant
cross-ratio. In particular, the CPWA of a scalar  and of a second-rank tensor
have the following leading terms \bea \label{sc} {\cal H}_{\Delta,
0}(x_1,x_2,x_3,x_4)&=&\frac{1}{x_{12}^4x_{34}^4} v^{\frac{\Delta}{2}} \Biggl(
1+\frac{\Delta}{4}Y +\frac{\Delta^3}{16(\Delta -1)(\Delta+1)}v
+\cdots\Biggl) \\
\label{ten} {\cal H}_{\Delta, 2}(x_1,x_2,x_3,x_4)&=&\frac{1}{x_{12}^4x_{34}^4}
v^{(\Delta-2)/2}\Biggl( \frac{1}{4}Y^2-\frac{1}{4}v \cdots\Biggl) \eea Assuming
that the dimension $\Delta$ takes the  form $\Delta=\Delta_0+\gamma$, where
$\gamma$ is the one-loop anomalous dimension, we see that the non-analytic term
$v^{\Delta/2}$ gives rise to the logarithmic terms of perturbation theory: \bea
v^{\gamma/2}=1+1/2\gamma \ln v +1/8 \gamma^2 \ln^2 v+... \eea Thus, comparing
the analytic terms in (\ref{sc}) and (\ref{ten}) against the ones in a concrete
four-point function allows one to identify the canonical dimensions and spins
of the contributing operators, while the logarithmic terms store the
information about their anomalous dimensions. Let us see how this works for our
operators $\Sigma_i$ of canonical dimension $\Delta_0=4$.

The operator with the lowest canonical dimension contributing to the OPE of two
operators $O_{\bf 20}^{LM}$ is the Konishi scalar $K_{s}$. Among the operators
with $\Delta_0=4$ one finds the stress-energy tensor $T$, the Konishi tensor
$K_{t}$, a new superconformal primary operator $\Xi$ and finally the operators
$\Sigma_i$ \cite{Arutyunov:2000im}. The operators $K_s$, $K_t$ and $\Xi$ are
from the $K$-class discussed in the Introduction and the stress-energy tensor
is protected. Thus, schematically we may write \bea \label{OPEld} O_{\bf
20}^{LM}O_{\bf 20}^{LM} = K_s+K_t+\Xi+T+\sum_{i=1}^4 \Sigma_i+\mbox{higher
dim}\, . \eea

Our quasiprimary operators $\Sigma_i$ are Lorentz scalars with $\Delta_0=4$
and, as is clear from (\ref{sc}), the lowest-power monomial occurring in their
CPWAs is $v^2$. Hence, looking for the coefficients of $v^2$, $v^2\ln v$ and
$v^2\ln^2v$ allows us to read off some information about their normalization
coefficients $A_i$ and anomalous dimensions $\gamma_i$. One should bear in
mind, however, that in principle the coefficients of the above mentioned
structures receive contributions from {\it all} operators participating in
(\ref{OPEld}). Thus, expanding Eq. (\ref{GE}) in powers of $\gamma$ up to the
second order, we find (only the relevant terms are indicated) \bea \nonumber
F&=&\frac{v^2}{x_{12}^4x_{34}^4}\Biggl[ \left(\sum_{i=1}^4 A_i
+\frac{1}{6}A_{K_{s}}
-\frac{1}{4}A_{K_{t}}-\frac{1}{4}A_{\Xi}-\frac{1}{4}A_{T}\right)\\
\label{4ptabs}
~~~&+&\frac{1}{2}\ln v \left(\sum_{i=1}^4 A_i\gamma_i
+\frac{1}{6}A_{K_s}\gamma_{K_{s}}
-\frac{1}{4}A_{K_{t}}\gamma_{K_{t}}-\frac{1}{4}A_{\Xi}\gamma_{\Xi}\right) \\
\nonumber~~~&+&\frac{1}{8}\ln^2v \left(\sum_{i=1}^4 A_i\gamma_i^2
+\frac{1}{6}A_{K_s}\gamma_{K_s}^2
-\frac{1}{4}A_{K_t}\gamma_{K_t}^2-\frac{1}{4}A_{\Xi}\gamma_{\Xi}^2\right)
\Biggr] \, , \eea
where we have also taken into account that the stress-energy tensor
is protected.

Since in the large $N$ limit the normalization coefficients and anomalous
dimensions of all operators in (\ref{OPEld}) except $\Sigma_i$ are already
known (see e.g. \cite{Arutyunov:2001mh}): \bea \label{coeff}
A_{K_{s}}=\frac{4}{3N^2}\, , ~~~~ A_{K_{t}}=\frac{16}{63N^2}\, , ~~~~
A_{\Xi}=\frac{16}{35N^2}\, ,~~~~A_T=\frac{8}{45N^2} , \eea and \bea
\label{anomK} \gamma_{K_s}=\gamma_{K_t}=3\lambda\, , ~~~~~
\gamma_{\Xi}=\frac{25}{6} \lambda\,  \eea one can substitute these
quantities\footnote{One can see that both combinations $\frac{1}{6}A_{K_{s}}
-\frac{1}{4}A_{K_{t}}-\frac{1}{4}A_{\Xi}-\frac{1}{4}A_{T}$ and
$\frac{1}{6}A_{K_s}\gamma_{K_{s}}
-\frac{1}{4}A_{K_{t}}\gamma_{K_{t}}-\frac{1}{4}A_{\Xi}\gamma_{\Xi}$ in fact
vanish.} in (\ref{4ptabs}).

The CPWA expansion (leading in $1/N^2$) of the four point function of the
operators $O_{\bf 20}^{LM}$ up to two loops was constructed in Ref.
\cite{Arutyunov:2001mh}. Thus, the part of this four-point function relevant to
our analysis can be extracted from Ref. \cite{Arutyunov:2001mh} and it reads
\bea \label{4ptaeps}
F=\frac{v^2}{x_{12}^4x_{14}^4}\Biggl[\frac{1}{10}\left(1+\frac{2}{3N^2}\right)-
\frac{1}{5N^2} \lambda \ln v+ \frac{7}{45N^2}\lambda^2\ln^2v \Biggr] \, .\eea
Finally, comparing eq. (\ref{4ptabs}) (with (\ref{coeff}) and (\ref{anomK})
inserted)  and eq. (\ref{4ptaeps}), we obtain the following set of equations
for the normalization constants and the one-loop anomalous dimensions of the
operators $\Sigma_i$ in  leading order in $1/N^2$: \bea \label{OPEcheck}
\sum_{i=1}^4 A_i \equiv g^2_0 =\frac{1}{10}\left(1+\frac{2}{3N^2}\right)  \, ,
~~~~ \sum_{i=1}^4 A_i\gamma_i=-\frac{2}{5}\frac{\lambda}{N^2} \, ,~~~~
\label{dtrs} \sum_{i=1}^4 A_i\gamma_i^2=\frac{9}{5} \frac{\lambda^2}{N^2} \, .
\eea It remains to note that for canonically normalized operators $\Sigma_i$
the coefficients $A_i$ coincide with the square of the normalization
coefficient of their three-point function with two $O_{\bf 20}$. As a
consequence, we obtain the consistency conditions (\ref{cons1})--(\ref{cons3})
on the diagonalization coefficients $a_i^2= g^{-2}_0 A_i$ in (\ref{Ofree}).
They were used in Section 5 to verify the agreement of our findings with the
OPE structure implied by one- and two-loop four-point functions of CPOs.

\section{Diagonalization procedure}

The four operators defined in (\ref{mix9}) are not orthonormal at tree level.
In fact, by computing their two-point functions (for simplicity we are now
restricting our attention to the lowest components) we find \be \langle {\cal
A}_i(x) {\cal A}_j(0)\rangle_0 = \frac{1}{(4\pi^2)^4}\frac{C_{ij}}{(x^2)^4} \ee
where \be \label{treematrixapp} C_{ij} = 3(N^2-1) \left( \matrix{ \frac{7 N^2 +
2}{2} & N^2 + 6 & \frac{7N^2 - 8}{N} & \frac{9N^2 - 16}{2N}\cr N^2 + 6 & 2(3
N^2 - 2) & \frac{2(N^2 - 4)}{N} & \frac{7N^2 - 8}{N}\cr \frac{7N^2 - 8}{N} &
\frac{2(N^2 - 4)}{N} & \frac{3N^4 - 8N^2 + 24}{N^2} & \frac{N^4 - 16N^2 +
48}{2N^2}\cr \frac{9N^2 - 16}{2N} & \frac{7N^2 - 8}{N} & \frac{N^4 - 16N^2 +
48}{2N^2} & \frac{7N^4 - 32N^2 + 96}{4N^2} } \right) \ee The diagonal $2 \times
2$ matrices are leading order in $N$ plus corrections order $\frac{1}{N^2}$ and
$\frac{1}{N^4}$, whereas the off--diagonal matrices contain only subleading
contributions order $\frac{1}{N}$ and $\frac{1}{N^3}$. Therefore, at leading
order in $N$ the double trace operators ${\cal A}_1$, ${\cal A}_2$ do not mix
with the single traces ${\cal A}_3$ and ${\cal A}_4$ and a block
diagonalization allows to easily obtain an orthogonal basis. A possible choice
might be $({\cal A}_1 - \frac{1}{6} {\cal A}_2)$, ${\cal A}_2$, $({\cal A}_3
-\frac{2}{7} {\cal A}_4)$ and ${\cal A}_4$.

At order $\frac{1}{N^2}$, keeping contributions from the matrix
(\ref{treematrixapp}) up to this order, we can define the
orthonormal basis
\bea {\cal O}_1 &=& \frac{1}{\sqrt{10}N^2}\left( 1 +
\frac{2}{3N^2}\right)\left[{\cal A}_1 -\frac{1}{6}{\cal A}_2 \right]
\non\\
{\cal O}_2 &=& \frac{1}{\sqrt{18}N^2}\left( 1 +
\frac{217}{84N^2}\right)\left[{\cal A}_2 -\frac{1}{N} \left( \frac{3}{2} {\cal
A}_3 + 3 {\cal A}_4 \right) + \frac{4}{N^2}\left({\cal A}_1 -\frac{1}{6}{\cal
A}_2 \right)
\right]\non\\
{\cal O}_3 &=& \sqrt{\frac{7}{60}}\frac{1}{N^2}\left( 1 + \frac{67}{28N^2}
\right) \left[ {\cal A}_3 -\frac{2}{7}{\cal A}_4 +\frac{1}{N}\left( {\cal A}_2
-\frac{12}{7}{\cal A}_1
\right)\right]\non\\
{\cal O}_4 &=& \frac{2}{\sqrt{21}N^2}\left( 1 +
\frac{61}{14N^2}\right)\left[{\cal A}_4 - \frac{1}{N} {\cal A}_1
+\frac{5}{2N^2} \left( {\cal A}_3 -\frac{2}{7}{\cal A}_4 \right) \right]
\label{treeeigenvectors} \eea which satisfies \be \langle {\cal O}_i(x) {\cal
O}_j(0)\rangle_0 = \frac{1}{(4\pi^2)^4}\frac{1}{(x^2)^4} \left[\delta_{ij} +
O\left(\frac{1}{N^3}\right) \right] \ee This is the basis we are going to use
in order to solve the mixing problem at one-loop.

We now move to the one-loop calculation. As explain in the main text,
divergent contributions arise in all the two-point functions
between any pair of operators ${\cal A}_1,\ldots,{\cal A}_4$
given in (\ref{ginvop}). Using {\em Mathematica} we can then compute
\be
\langle {\cal A}_i (x){\cal A}_j (0) \rangle_1 = \frac{1}{\e}
\frac{\o_{ij}}{(4\pi^2)^4} \frac{1}{(x^2)^4}
\ee
where
\be
\label{loopmatrixapp}
\o = -\frac{3}{2}\lambda(N^2-1)
\left(
\matrix{
-2N^2 + 13 & -6(2N^2 + 7) & \frac{21N^2 + 16}{N} & -\frac{53N^2 - 32}{2N}\cr
-6(2N^2 + 7) & -12(6N^2 + 1) & \frac{6(N^2 + 16)}{N} & -\frac{3(33N^2 - 32)}
{N}\cr
\frac{21N^2 + 16}{N} & \frac{6(N^2 + 16)}{N} & -\frac{11N^4 - 96N^2 + 128}
{N^2} & \frac{3N^4 + 112N^2 - 256}{2N^2}\cr
-\frac{53N^2 - 32}{2N} & -\frac{3(33N^2 - 32)}{N} & \frac{3N^4 + 112N^2 - 256}
{2N^2} & -\frac{59N^4 - 64N^2 + 512}{4N^2}
}
\right)
\ee
is the anomalous dimensions matrix. It is not diagonal since there is mixing
among the operators. We then face the problem to solve the mixing
perturbatively in $\frac{1}{N}$, for large $N$. We will determine the anomalous
dimensions up to order $\frac{1}{N^2}$ and the corresponding quasiprimary
operators up to $\frac{1}{N}$.

First of all, we have to diagonalize the matrix of the two--point functions at
tree level. The new basis, up to the order we are interested in, is given in
(\ref{treeeigenvectors}). On this basis the matrix $\o$ is given by \bea
\label{admatrix} && \o = \lambda \left[ \left( \matrix{ 0 & 0 & 0 & 0 \cr 0 & 6
& 0 & 0 \cr 0 & 0 & \frac{16}{7} & \frac{4\sqrt{5}}{7} \cr 0 & 0 &
\frac{4\sqrt{5}}{7} & \frac{59}{14} } \right) + \frac{1}{N}\sqrt{\frac{6}{7}}
\left( \matrix{ 0 & 0 & -4 & -\sqrt{5} \cr 0 & 0 & \sqrt{5} & -\frac{15}{2} \cr
-4 & \sqrt{5} & 0 & 0 \cr -\sqrt{5} & -\frac{15}{2} & 0 & 0 } \right) \right.
\non\\
&&~~~~~~\non\\
&&~~~~~~~~~\left. +\frac{1}{N^2} \left( \matrix{ -4 & 2\sqrt{5} & 0 & 0 \cr
2\sqrt{5} & -10 & 0 & 0 \cr 0 & 0 & \frac{431}{49} & -\frac{215\sqrt{5}}{98}
\cr 0 & 0 & -\frac{215\sqrt{5}}{98} & \frac{255}{49}} \right) \right] +
O\left(\frac{1}{N^3}\right) \eea from which it is clear that we have a zero
eigenvalue at leading order in $N$.

Again by using {\em Mathematica} we have computed its eigenvalues up
to order $\frac{1}{N^2}$
\bea
\label{eigenv1}
&& \gamma_1= -10\frac{\lambda}{N^2} \qquad , \qquad
\gamma_2= \left(6 +
  \frac{20}{N^2}\right)
{\lambda} \nonumber \\
&& \gamma_3= \left[\frac{13 + \sqrt{41}}{4} -
  \frac{5}{41N^2}(41+19\sqrt{41})\right]
{\lambda} \nonumber \\
&& \gamma_4=\left[\frac{13 - \sqrt{41}}{4} -
  \frac{5}{41N^2}(41-19\sqrt{41})\right]
{\lambda} \eea
and the corresponding eigenvectors
\bea && \S_1=
\left(1-\frac{21}{16N^2}\right){\cal O}_1 + \frac{\sqrt{42}}{4N} {\cal O}_3
-\frac{7\sqrt{5}}{12N^2}{\cal O}_2 +
O\left(\frac{1}{N^3}\right)\nonumber\\
&&\S_2= \left(1-\frac{45}{4N^2}\right){\cal O}_2
-\frac{1}{N}\sqrt{\frac{3}{14}} \left(\sqrt{5} {\cal O}_3 +10 {\cal
O}_4\right)+\frac{4\sqrt{5}}{3N^2}{\cal O}_1 +
O\left(\frac{1}{N^3}\right)  \nonumber \\
&& \S_3= \sqrt{\frac{287-27\sqrt{41}}{574}} \left[{\cal O}_3 +
\frac{27+7\sqrt{41}}{16\sqrt{5}} {\cal O}_4 -\frac{\sqrt{42}}{4N}\left({\cal
O}_1 - \sqrt{5}\frac{\sqrt{41}+5}{4}{\cal O}_2
\right)\right] + O\left(\frac{1}{N^2}\right) \nonumber \\
&& \S_4= \sqrt{\frac{287+27\sqrt{41}}{574}} \left[{\cal O}_3 +
\frac{27-7\sqrt{41}}{16\sqrt{5}} {\cal O}_4 -\frac{\sqrt{42}}{4N}\left({\cal
O}_1 + \sqrt{5}\frac{\sqrt{41}-5}{4}
{\cal O}_2 \right)\right] + O\left(\frac{1}{N^2}\right) \nonumber\\
&&\nonumber\\
&& \label{eigenvectors1} \eea These are one-loop quasiprimary operators with
anomalous dimensions $\g_i$. They are uniquely defined, because the anomalous
dimensions are all different (cf.. the general discussion in Section 3.2). At
this order these operators satisfy \be \langle \S_i(x) \S_j(0) \rangle_{0+1}
\sim \frac{1}{(4\pi^2)^4} \frac{1}{(x^2)^4} \d_{ij}\left[ 1 - \g_i \ln{(x^2
\mu^2)} + O\left( \frac{1}{N^2} \right) \right] \ee up to a finite
renormalization constant which we have not considered.

Inverting eq. (\ref{eigenvectors1}) we can express the operator ${\cal O}_1$ in
terms of pure states \be \label{O1split} {\cal O}_1 = \S_1 -
\frac{1}{N}\left[\a_3\S_3 + \a_4\S_4\right]
+\frac{1}{N^2}\left[-\frac{21}{16}\S_1 + \frac{3\sqrt{5}}{4}\S_2 \right] +
O\left(\frac{1}{N^3}\right) \ee where \bea \label{agamma} &&\a_3 =
-\frac{27-7\sqrt{41}}{64}\sqrt{42\left( \frac{1}{10}+
\frac{27}{70\sqrt{41}}\right)} \nonumber \\
&&\a_4 = \frac{27+7\sqrt{41}}{64}\sqrt{42\left(\frac{1}{10}-
\frac{27}{70\sqrt{41}}\right)}
\eea

Similarly, from (\ref{treeeigenvectors}) and (\ref{eigenvectors1})
we can determine the (normalized) operator ${\cal A}_2$, which we call
${\cal K}^2$,
in terms of pure states
\be \label{Konishisquared}
{{\cal K}}^2 \equiv\frac{1}{\sqrt{18}N^2}\left(1+\frac{5}{6N^2}\right)
{\cal A}_2
= \S_2 + \frac{1}{N}\left[\b_3\S_3 + \b_4\S_4\right]
+\frac{1}{N^2}\left[\frac{5\sqrt{5}}{6}\S_1 -
\frac{162}{7} \S_2 \right] + O\left(\frac{1}{N^3}\right) \, ,
\ee
where
\bea \label{bgamma}
&&\b_3 =
\frac{287+85\sqrt{41}}{4592} \sqrt{861+81\sqrt{41}} \nonumber \\
&&\b_4 =
\frac{287-85\sqrt{41}}{4592} \sqrt{861-81\sqrt{41}}
\eea

\end{document}